\pdfoutput=1
\documentclass[%
aps,
reprint,
superscriptaddress,
nofootinbib,
amsmath,
amssymb,
dvipsnames
]{revtex4-2}

\usepackage{mathtools, nicefrac, physics}
\usepackage{array, multirow, tabularx, makecell, threeparttable}
\usepackage{graphicx, adjustbox, placeins}
\usepackage{enumitem, comment, ragged2e}
\usepackage[ruled, vlined, linesnumbered]{algorithm2e}

\usepackage[compat=0.3]{yquant}
\usepackage{quantikz}
\useyquantlanguage{groups}

\usepackage{hyperref}
\usepackage[capitalize]{cleveref}
\usepackage[caption=false]{subfig}

\hypersetup{
    colorlinks,
    linkcolor={blue!50!black},
    citecolor={blue!50!black},
    urlcolor={blue!80!black}
}
\Crefname{section}{Sec.}{Secs.}
\Crefname{subsection}{Sec.}{Secs.}
\crefname{appendix}{Appendix}{Appendices}
\Crefname{appendix}{Appendix}{Appendices}
\graphicspath{{figures/}}
\newcommand{\fiteqn}[1]{\resizebox{\hsize}{!}{$#1$}}

\begin{document}

\title{Layer codes as quantum memories: syndrome extraction, thresholds and idle robustness}

\newcommand{\qmaddress}{\affiliation{Quantum Motion, 9 Sterling Way, London N7 9HJ, United Kingdom}}
\newcommand{\oxddress}{\affiliation{Department of Materials, University of Oxford, Parks Road, Oxford OX1 3PH, United Kingdom}}
\newcommand{\mathinst}{\affiliation{Mathematical Institute, University of Oxford, Woodstock Road, Oxford OX2 6GG, United Kingdom}}

\newcommand{\oxengaddress}{\affiliation{Department of Engineering Science, University of Oxford, Parks Road, Oxford OX1 3PJ, United Kingdom}}

\newcommand{\icaddress}{\affiliation{Department of Computing, Imperial College London, 180 Queen's Gate, London SW7 2AZ, United Kingdom}}

\author{Zhu Sun}
\email{zhu.sun@exeter.ox.ac.uk}
\oxddress
\mathinst
\qmaddress

\author{Zhenyu Cai}
\email{z.cai1@imperial.ac.uk}
\icaddress
\oxengaddress
\qmaddress

\begin{abstract}
Layer codes are three-dimensional local CSS codes of check weight at most six, obtained by quasi-concatenating a CSS input code with unrotated surface code patches. What is established about them describes the code, not the circuit that measures it. In this work, we evaluate them as active memories under circuit-level noise. Doing so first requires a syndrome extraction circuit, and the junction stabilizers that couple the patches rule out borrowing a surface code patch's CNOT schedule. We present a decoder-free scheduler that constrains hook propagation first and compresses depth afterwards, generate circuits for eighteen layer codes across four input codes, and compare them with surface codes of the same distance, simulated and decoded in the same way. The circuits lose no distance to hook errors on $[[4,2,2]]$-input codes, and reach a lower logical error rate than those of a general-purpose scheduler for CSS codes at roughly half its CNOT depth per round. Our primary family built from a $[[4,2,2]]$ input code reaches a circuit-level threshold of $5.06(7)\times10^{-3}$. The layer code outperforms the surface code in terms of idle robustness: at the same distance it tolerates an interval between syndrome extraction rounds $1.7$--$3.0$ times longer than the rotated surface code. This comes at the cost of $1.7$--$5.0$ times more operations per unit time and approximately five times more physical qubits, both per logical qubit. Finally, we model the cost of the beyond-2D cross-plane CNOTs that the construction demands, and find that their speed matters far more than their fidelity.
\end{abstract}

\maketitle

\section{Introduction}
\label{sec:intro}

Quantum error correction protects logical information by measuring the checks of a code over and over. What that costs in physical qubits, in gates and in time sets the overhead of a fault-tolerant architecture. And which code is worth the cost depends on what the hardware can measure, not only on the code's parameters. The surface code is one of the most promising candidates for the first generation of quantum computers, because it has a high threshold and its checks act on nearest neighbours of a two-dimensional lattice, which is what most planar hardware supplies. Yet the price is the encoding rate: a patch stores only one logical qubit, and its size scales quadratically with the code distance. Quantum low-density parity-check (qLDPC) codes of constant rate and growing distance avoid that price~\cite{panteleev2022almost,leverrier2022quantum,bravyi2024high}, but their checks usually demand complicated connectivity. Between the two lie codes that are still geometrically local, just in more than two dimensions. The trade-off between the achievable rate and distance is described by the Bravyi--Poulin--Terhal (BPT) bound, which becomes less restrictive with increasing spatial dimension~\cite{PhysRevLett.104.050503}, allowing higher-dimensional codes to evade constraints imposed by the plane.

Layer codes~\cite{layercodes2309} sit in that middle ground in three dimensions. The construction quasi-concatenates a CSS input code with unrotated surface code patches and returns a code that is local in 3D with checks of weight at most six; with good qLDPC inputs the family saturates the BPT bound, proved with quantum Tanner codes~\cite{leverrier2022quantum} as the input family. As passive memories coupled to a finite-temperature bath they are also partially self-correcting, their storage time growing with system size up to a temperature-dependent cutoff, with the syndrome there read noiselessly~\cite{layercodepy2510,williamson2025partialselfcorrectionlayercodes}. To our knowledge, no previous work has explored what happens when that syndrome is extracted repeatedly by a noisy circuit, which is how a memory is actually run.

The missing piece is the syndrome extraction circuit. Where the planes of a layer code meet, the construction places junction stabilizers of weight five and six whose support spans several planes, so neither a surface code patch's CNOT schedule nor its hook-error analysis carries over, and a schedule chosen carelessly can cost circuit distance.

This paper builds that circuit and uses it to measure layer codes as circuit-level memories. We design a syndrome extraction scheduler for them and generate circuits for eighteen layer codes across four input codes; we decode globally with BP+LSD on the raw detector error model; and we run the same experiments on unrotated and rotated surface codes of the same distance, with every convention held fixed across codes.

Our contributions are the following.
\begin{itemize}
    \item \textbf{A syndrome extraction scheduler for layer codes} (\cref{sec:sec-design}). The scheduler first constrains the CNOT ordering within each stabilizer using a geometric rule that limits hook propagation, then assigns each CNOT to a time step while achieving provably minimal depth subject to detector determinism and the permitted orderings. In the circuits we simulate, this eliminates virtually all ancilla idle time between reset and measurement. For layer codes constructed from the $[[4,2,2]]$ input code, our circuits lose no circuit distance to hook errors and achieve lower logical error rates than those produced by a general-purpose CSS-code scheduler~\cite{strikis2026highperformance}, at approximately half the CNOT depth per round.
    \item \textbf{Circuit-level thresholds and the origin of the gap} (\cref{sec:memory-ler}). The layer code has a lower threshold than the surface code, and we show that this gap arises primarily from the greater depth of its syndrome-extraction rounds rather than the quality of its hook structure. Our primary family, constructed from a $[[4,2,2]]$ input code, achieves thresholds of $p_{\mathrm{th}}=4.99(6)\times10^{-3}$ for X memory and $5.06(7)\times10^{-3}$ for Z memory, compared with $7.71(9)\times10^{-3}$ and $7.73(14)\times10^{-3}$ for the rotated surface code, and $8.02(6)\times10^{-3}$ and $8.29(6)\times10^{-3}$ for the unrotated surface code, whose patches serve as the constituent patches of the layer code.
    \item \textbf{Idle robustness} (\cref{sec:idle}). We quantify how infrequently error correction needs to be performed by considering a logical state stored over repeated periods, each consisting of a few rounds of syndrome extraction followed by an idle interval. We define the \emph{sustainable idle interval} as the idle duration at which a single period randomizes the stored logical state with a probability of $1\%$. At equal code distance, the layer code sustains idle intervals $1.7$--$3.0$ times longer than the rotated surface code, at the cost of $1.7$--$5.0$ times more two-qubit gates, measurements, and noisy locations per logical qubit per unit time. It also requires approximately five times as many physical qubits per logical qubit.
    \item \textbf{The cost of the cross-plane gates} (\cref{sec:hardware}). For layer codes, the speed of the required beyond-2D cross-plane gates matters more than their fidelity. Increasing their error rate tenfold raises the logical error rate by only about a factor of two, whereas doubling their duration increases it by a factor of five to seventeen. We also examine how much this speed requirement depends on scheduling.
\end{itemize}

\Cref{sec:background} fixes the conventions used throughout, and \cref{sec:discussion} collects the routes that did not work, the limits of what we measured, and directions for future work.

\section{Background}
\label{sec:background}
We introduce the relevant background and conventions required for understanding this work. Basic QEC knowledge is assumed.

\subsection{Layer Codes}
\begin{figure*}[t]
    \centering
    \includegraphics[height=0.36\textwidth]{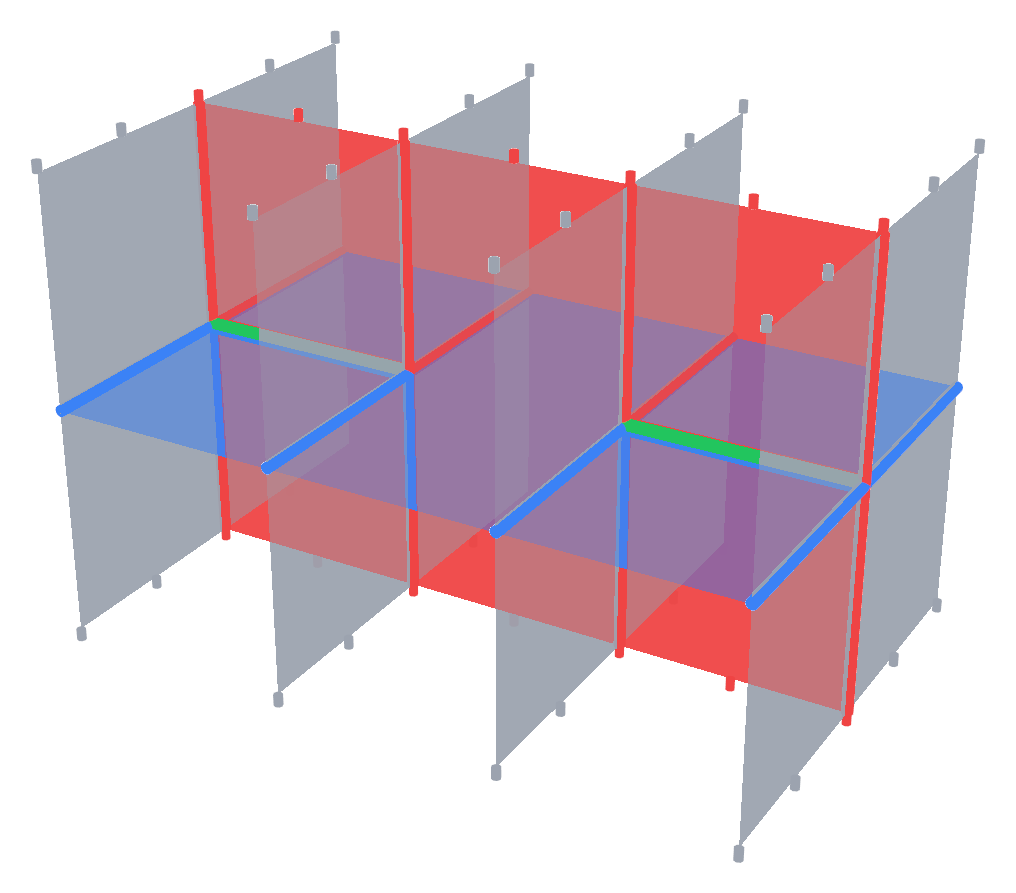}
    \hfill
    \includegraphics[height=0.36\textwidth]{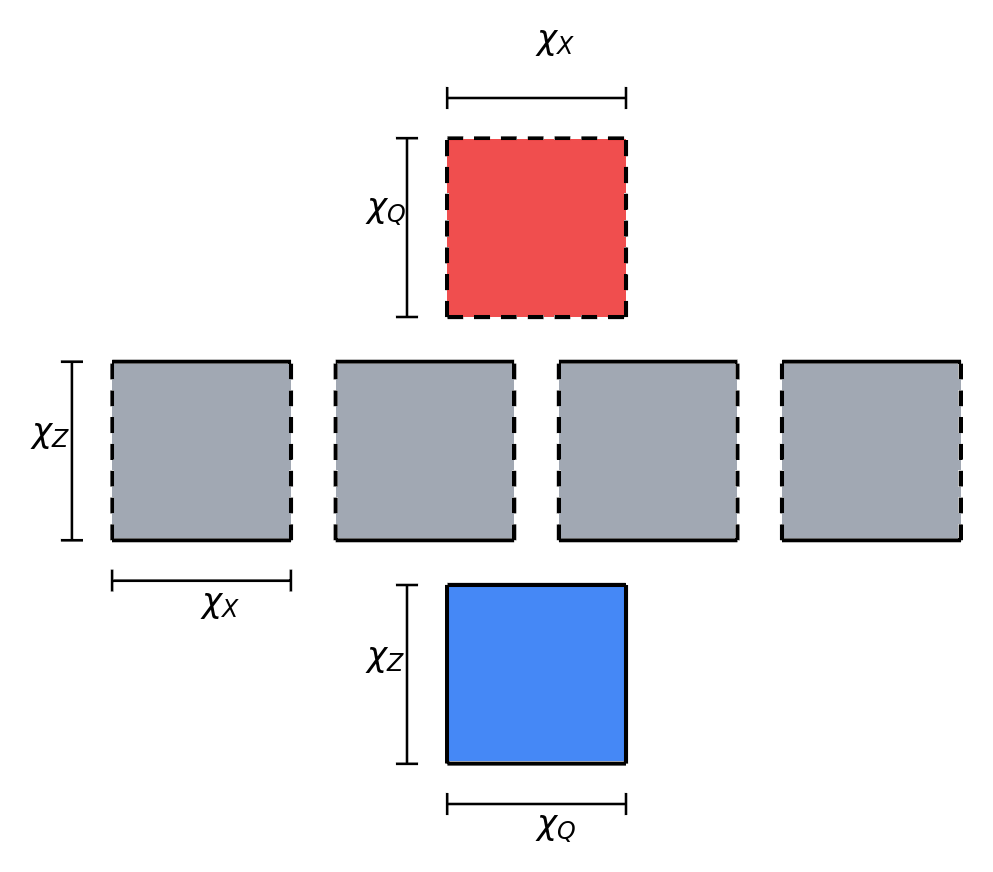}
    \caption{The layer code with input code $[[4,2,2]]$ in its two constructions. Left: the WB24 3D local code -- grey data planes interleaved with the blue X-check and red Z-check planes; the coloured lines mark the one-dimensional junctions where planes intersect. Plain and bumped plane edges denote smooth and rough surface code boundaries, respectively. Right: the YBW26 form of the same code (shown with $\chi=(4,4,4)$) -- a 2D array of surface code patches, one grey data patch per input qubit and one blue (red) check patch per X-type (Z-type) input check, with patch dimensions set by $\chi=(\chi_X,\chi_Q,\chi_Z)$ as annotated. Solid and dashed patch boundaries denote smooth and rough surface code boundaries, respectively.}
    \label{fig:layer-codes}
\end{figure*}

\begin{itemize}
    \item \textbf{Layer codes.} A layer code~\cite{layercodes2309} is a 3D local code obtained by quasi-concatenating a CSS input code with unrotated surface codes. Rather than a single code family, the construction is a recipe: for an input CSS code with parameters $[[n,k,d]]$, it returns a 3D local CSS code with parameters $[[N, K, D]]=[[\Theta(nn_X n_Z ), k, \Omega(\frac{1}{w} d\min(n_X,n_Z ))]]$ and check weight at most 6, where $n_X$ and $n_Z$ denote the numbers of X- and Z-type checks of the input code and $w$ its maximum check weight. The qubits and checks of the output layer code are organized into parallel planes: data planes hosting surface code patches, interleaved with X-check and Z-check planes derived from the input code's parity checks. An example of a layer code with input code $[[4,2,2]]$ is shown in the left panel of \cref{fig:layer-codes}. In the following discussion, we refer to the construction conventions following ref.~\cite{layercodes2309} as WB24.

    Layer codes have attracted considerable attention because, with good quantum low-density parity-check (qLDPC) input codes, they saturate the 3D Bravyi--Poulin--Terhal (BPT) bound \cite{PhysRevLett.104.050503}, which characterizes the optimal trade-off between encoding rate and code distance for geometrically local quantum codes. They also exhibit partial self-correction as passive quantum memories under finite-temperature Hamiltonian dynamics, where the memory lifetime increases with system size up to a temperature-dependent cutoff \cite{layercodepy2510,williamson2025partialselfcorrectionlayercodes}. These results, however, concern passive quantum memories rather than the active circuit-level fault-tolerance setting considered in this work, where errors arise from noisy physical operations and are corrected through repeated syndrome extraction and decoding.

    \item \textbf{Junction stabilizers.} Where check planes meet data planes (where the grey planes intersect the red and blue planes in the left panel of \cref{fig:layer-codes}), the construction produces weight-5/6 stabilizers whose support spans multiple planes. These junction stabilizers are the structural origin of the syndrome extraction scheduling problem addressed in \cref{sec:sec-design}. Away from the junctions, the stabilizers are those of the surface code patches, of weight 4 in the bulk and 2 or 3 at patch boundaries.
    \item \textbf{YBW26 convention.} Ref.~\cite{yuan2026quantumweightreductionlayer} reformulated the construction of layer codes with algebraic homology and introduced more flexibility in dimensions. This new convention characterizes each layer code by the parameter $\chi=(\chi_X,\chi_Q,\chi_Z)$: X-check, data, and Z-check patches have dimensions $\chi_Q\times\chi_Z$, $\chi_X\times\chi_Z$, and $\chi_X\times\chi_Q$, respectively (\cref{fig:layer-codes}), so $\chi$ also acts as an independent distance parameter. Moreover, it flattens the construction from a 3D local code into a 2D array of surface code patches with nonlocal inter-patch couplings; an example with input code $[[4,2,2]]$ is shown in the right panel of \cref{fig:layer-codes}. Intuitively, it unpacks the 3D intersecting layers onto 2D, and arranges the patches into the shape of a Tanner graph of the input code.

    We refer to this convention as YBW26 and it is the default construction convention. We will use, e.g., ``the $[[4,2,2]]$ family'' to denote the YBW26 layer codes with $[[4,2,2]]$ input code.

    \item \textbf{Distance pair $(d_X, d_Z)$.} The layer code construction is not dual-covariant: $d_X \neq d_Z$ is generic, even for self-dual input codes. We therefore report layer codes with both distances, $(d_X, d_Z)$, and will only use ``distance $d$'' when $d_X = d_Z$.

\end{itemize}

\subsection{Circuits and Scheduling}
All codes in this work are CSS, so the definitions below assume this structure.

\begin{itemize}
    \item \textbf{Syndrome-extraction circuit (SEC).} An SEC implements one round of stabilizer measurements using ancilla qubits. For each X-type (Z-type) stabilizer, the ancilla is initialized in the $\ket{+}$ ($\ket{0}$) state, interacts with the corresponding data qubits through CNOT gates, and is finally measured in the X (Z) basis. Throughout this work, ancilla preparation, single- and two-qubit gates, measurements, and idle operations each occupy one circuit tick, except that in the SEC an ancilla's measurement and its reset for the next round are fitted into one tick.

    Since each qubit can participate in at most one two-qubit gate per tick, the CNOT schedule corresponds to an edge colouring of the qubit--stabilizer interaction graph. The number of colours determines the CNOT depth of one SEC round, while ancilla qubits remain idle during ticks in which their associated stabilizer is not being measured. These idle periods contribute noise and therefore form an important optimization target.

    \item \textbf{Hook errors.} A single circuit fault, typically on an ancilla qubit during syndrome extraction, can propagate through the subsequent CNOT gates of a stabilizer measurement and produce a correlated multi-qubit data error. Such hook errors may reduce the effective distance of the circuit below the code distance, making the CNOT ordering within each stabilizer an important scheduling design choice.

    \item \textbf{Detector error model (DEM).} A detector is a parity of measurement outcomes that is deterministic in the absence of faults. The DEM represents the noisy circuit as a collection of independent error mechanisms, each annotated with the detectors and logical observables it flips~\cite{gidney2021stim}. A mechanism that flips more than two detectors is called a hyperedge.
    \item \textbf{Circuit distance.} The circuit distance $d_{\mathrm{circ}}$ of an SEC is the minimum number of elementary error mechanisms that together flip a logical observable while flipping no detector. The circuit distance satisfies $d_{\mathrm{circ}}^{X}\le d_X$ and $d_{\mathrm{circ}}^{Z}\le d_Z$, with hook errors the primary mechanism that opens each gap. Standard estimates of $d_{\mathrm{circ}}$ are unavailable for layer codes: their DEMs contain hyperedges, on which graphlike shortest-error search is unreliable. We therefore determine $d_{\mathrm{circ}}$ by exhaustive search on the raw DEM (\cref{sec:sec-design}) using QEDist (version 0.1)~\cite{lee2026qedist}. Specifically, if the search establishes that no set of fewer than $w$ error mechanisms can flip a logical observable without flipping any detector, and an explicit set of $w$ mechanisms achieving this is identified, then $d_{\mathrm{circ}}=w$.

    The scheduler of ref.~\cite{strikis2026highperformance} instead ranks candidate schedules by a cheaper proxy, the minimum residual distance $\Delta_{\min}$. A mid-schedule ancilla fault propagates through the remaining CNOTs of the stabilizer and leaves a residual Pauli error on the data qubits. The residual distance is one plus the minimum weight of a logically nontrivial completion of that propagated error (Definition 1 in ref.~\cite{strikis2026highperformance}), which is estimated with the BP+OSD decoder described in \cref{sec:decoding}. For a given schedule, $\Delta_{\min}$ is the minimum residual distance over all stabilizers and all fault positions in their CNOT orders.
    \item \textbf{Memory experiment.} An X memory (Z memory) experiment consists of preparing all data qubits in the X (Z) basis, $d_Z$ ($d_X$) rounds of SEC, and a final transversal readout in the memory basis. We emphasize that an X memory (Z memory) experiment requires $d_Z$ ($d_X$) rounds of SEC, since the measured logical observable is flipped only by logical operators of the opposite Pauli type. The memory experiment is our standard benchmark.
    \item \textbf{CSS-marginal detector convention.} Under this convention, mid-circuit detectors are placed only on stabilizers associated with the measured memory basis (e.g., X memory experiments include only X-stabilizer detectors). This leaves the physical circuit unchanged but modifies the DEM seen by the decoder, leading to logical error rates that may differ by a factor of several. Consequently, all comparisons across code families and SEC schedules are performed using a consistent detector convention. Unless stated otherwise, all results reported in this work use the CSS-marginal convention.
    \item \textbf{Logical observable representatives.} A logical operator is an equivalence class modulo stabilizers, whereas the final readout of a memory experiment measures a specific representative of that class. Different circuit schedulers may therefore use different logical observable representatives. Since the measured logical observable depends on this choice, comparisons across schedulers must use aligned representatives.

\end{itemize}

\subsection{Noise Models}
The noise models used in our benchmarks are introduced below.
\begin{itemize}
    \item \textbf{Code capacity noise model.} Only data qubits are affected by noise; syndrome extraction and measurements are perfect. It is usually used to evaluate an upper bound on the performance of a QEC code.
    \item \textbf{Phenomenological noise model.} Data qubits are affected by noise (as in the code-capacity noise model), and measurements are also faulty.
    \item \textbf{Circuit-level noise model.} Every operation (idling, physical one- and two-qubit gates, initialization, and measurement) is faulty and has an independent chance of inducing an error. This model is considered more realistic than the two above.
    \item \textbf{Depolarizing noise.}
    For error probability $p$, the single-qubit depolarizing channel is defined as
    \begin{equation}
        \text{DEP1}(p,\rho) = (1-p)\rho + \frac{p}{3}(X\rho X+Y\rho Y+Z\rho Z).
    \end{equation}
    The two-qubit depolarizing channel is defined as
    \begin{equation}
        \text{DEP2}(p,\rho) = (1-p)\rho + \frac{p}{15}\sum_{P\in \mathcal{P}} P\rho P,
    \end{equation}
    where $\mathcal{P} = \{I,X,Y,Z\}^{\otimes 2} \setminus \{I \otimes I\}$. This is the most commonly used circuit-level noise model and the one we adopt. Throughout this paper, all error rates are fixed by a single physical error parameter $p$.
    \item \textbf{Per-round idle (depth-agnostic).} A circuit-level noise model convention. At the beginning of each SEC round, one single-qubit depolarizing channel is applied to every qubit, data and ancilla alike, independent of the SEC depth; it is the depth-agnostic analogue of the convention of Stim's generated circuits~\cite{gidney2021stim}.
    \item \textbf{Per-tick idle (depth-aware).} A circuit-level noise model convention. During SEC, a single-qubit depolarizing channel is applied on every tick (time slice) to each idle qubit: a data qubit not acted on by a two-qubit gate, or an ancilla qubit not acted on by any operation. The circuit depth thus enters the logical error rate directly. This is the more physical model and our default convention. Still, we always state the idle model in use, as depth-mismatched comparisons can reverse verdicts between the two models.

\end{itemize}

\subsection{Decoding}
\label{sec:decoding}
\begin{itemize}
    \item \textbf{BP-based decoders.} BP-based decoders combine belief propagation, originally developed for decoding classical LDPC codes, with optional post-processing (e.g., OSD or LSD) to efficiently decode qLDPC codes using soft reliability information.

    We use BP+LSD~\cite{hillmann2024lsd} as our default decoder; BP+OSD~\cite{roffe2020decoding} is used as an accuracy reference on circuits small enough to afford it, such as the SEC noise-model benchmark of \cref{sec:sec-design}. We decode the raw DEM directly, preserving hyperedges rather than decomposing them into graphlike edges, and perform global decoding without plane decomposition. The per-plane matching decoders previously proposed for layer codes~\cite{layercodepy2510,williamson2025partialselfcorrectionlayercodes} rely on both decompositions and lose accuracy under circuit-level noise; \cref{sec:negative} quantifies this.
    \item \textbf{Logical error rate (LER).} The probability that the decoded logical state differs from the ideal one. Operationally, this corresponds to the probability that the decoded value of the logical observable differs from its ideal value. For input codes with $k>1$, each logical qubit carries its own observable, and a shot counts as a logical error if any of the $k$ observables is decoded incorrectly. We therefore report the LER normalized by its saturation value $1-2^{-k}$, i.e., $\overline{\mathrm{LER}}=\mathrm{LER}/(1-2^{-k})$.
    \item \textbf{Threshold vs. apparent threshold.} We distinguish the asymptotic threshold, estimated from a finite-size scaling collapse (used in \cref{sec:experiments}), from the apparent threshold, defined as the crossing point of finite-size LER curves (used in the noise-model benchmark of \cref{sec:sec-design}). The two estimators generally differ slightly when applied to the same dataset.
\end{itemize}

\section{SEC design}
\label{sec:sec-design}
Although the code distance sets the theoretical error-correction capability of a QEC code, its practical performance under circuit-level noise is determined by the syndrome-extraction circuit. For a fixed code, the SEC determines how closely the circuit distances $d_{\mathrm{circ}}^{X}$ and $d_{\mathrm{circ}}^{Z}$ approach their upper bounds $d_X$ and $d_Z$, making SEC design a central problem in fault-tolerant implementations.

Our SEC design is inspired by the method of ref.~\cite{strikis2026highperformance}, an SEC design tool for general CSS codes (referred to as LRC in the following). LRC first generates colouring candidates at a fixed low depth, ranks them by proxies for hook-error harm (the minimum residual distance $\Delta_{\min}$, with ancilla idle count and the residual-distance profile as tie-breakers), then picks the top candidate. Ranking the candidates this way requires a BP+OSD evaluation of every residual, which is where LRC's construction time concentrates. We note that other schedulers for general CSS codes have been proposed that aim to minimize CNOT depth~\cite{zhang2026optimal,durazorocha2026localasc}, target minimum-weight logical errors in the circuit~\cite{viszlai2026prophunt}, or directly optimize the decoder's LER using Monte Carlo tree search~\cite{liu2026alphasyndrome} or reinforcement learning~\cite{ye2026reinforcement}. However, these approaches are beyond the scope of this work.

In comparison, our scheduler is decoder-free and splits into two phases. The first phase narrows each stabilizer's CNOT order. Its key ingredient is a geometric rule distilled from the hook-error metrics underlying LRC's ranking: the last CNOT of each stabilizer must act on a data qubit in its support that lies farthest, in lattice distance, from the least-weight of the $k$ cleaned logical representatives of the same type. This rule requires no BP+OSD evaluation, yet the orders it admits include the metric-best one for most stabilizers, and in a paired test, scheduling by the rule alone showed no detectable LER penalty relative to the metric-based ranking. The second phase is a constraint program. It assigns every CNOT a tick and simultaneously picks each stabilizer's order from the ones the rule admits. The depth it reaches is provably minimal, given detector determinism and that choice of orders. An optional part of the second phase then reduces ancilla idle exposure at that depth: each ancilla's reset and measurement are staggered to sit tight against its own CNOT chain, and the remaining mid-extraction idle ticks are minimized by an anytime constraint-programming step whose time budget is tunable. In the circuits we simulate this part removes essentially all ($100\%$ on 32 of the 36 circuits we built and above $93\%$ on the rest) of the ancilla idle ticks between reset and measurement, the only ones whose noise can propagate into the syndrome, which matters precisely under the depth-aware noise model of \cref{sec:background}. \Cref{app:sec-algorithm} states both phases as pseudocode.

\begin{figure*}[t]
    \centering
    \includegraphics[width=\textwidth]{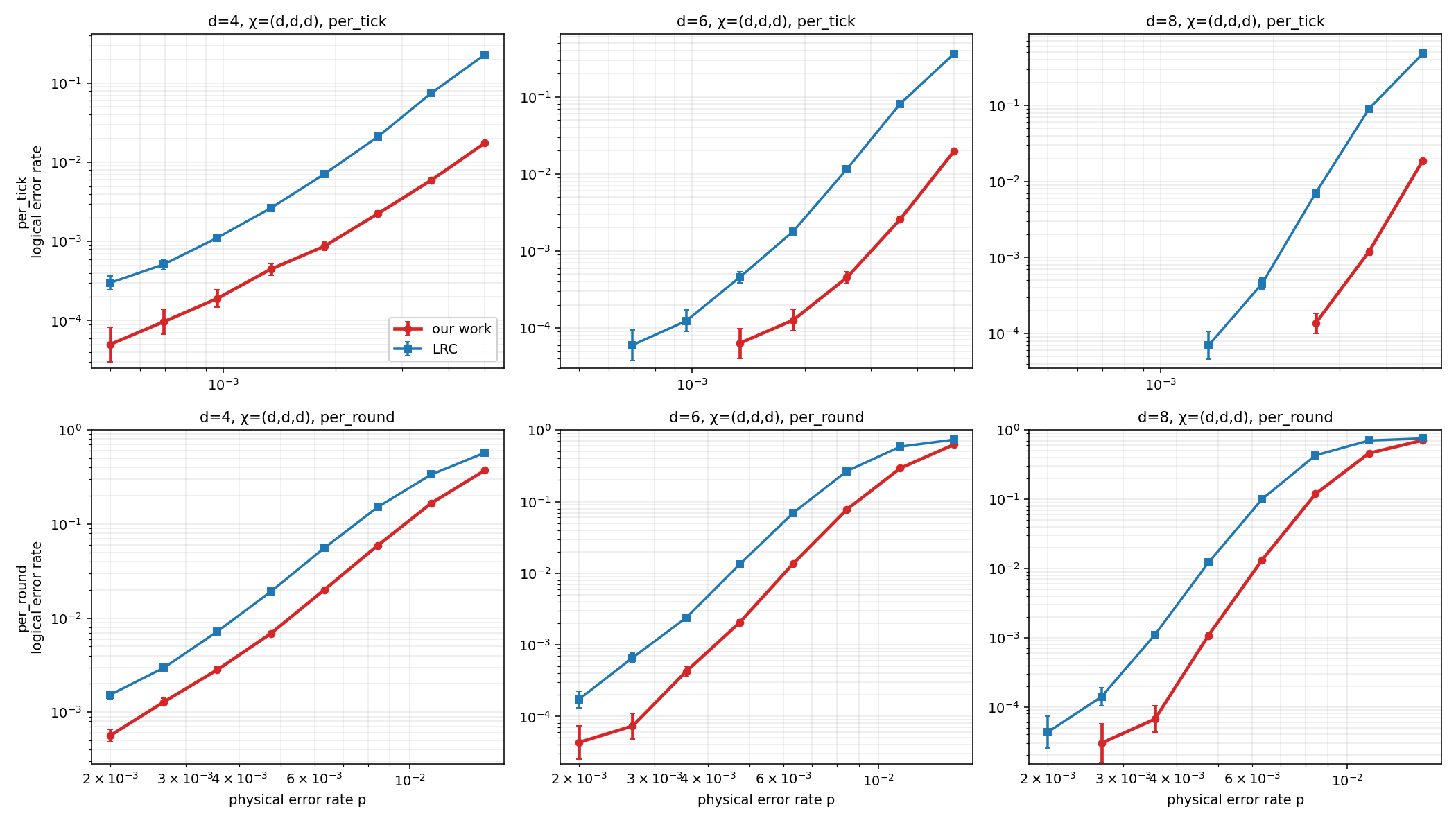}
    \caption{Logical error rate of our SEC (red) versus the LRC scheduler (blue) on the YBW26 $[[4,2,2]]$ layer code family, $\chi=(d,d,d)$, $d=4,6,8$ (columns): X-basis memory, $d$ rounds, mid-circuit detectors on memory-basis stabilizers only (CSS-marginal convention), decoded with a global BP+LSD decoder on the raw detector error model. Top row: depth-aware per-tick idle noise (default noise model); bottom row: depth-agnostic per-round idle noise. Error bars are Wilson 95\% intervals. Points with fewer than five observed logical errors are omitted. The two circuits are sampled independently (unpaired comparison). Each circuit uses its scheduler's own (equivalent) logical representative; however, replacing LRC's representative with ours leaves its logical error rate statistically unchanged.}
    \label{fig:sec-ler-lrc}
\end{figure*}

We note the two different design philosophies: LRC fixes a low circuit depth first and optimizes its hook metrics within that fixed depth, whereas we constrain hook propagation first and compress depth afterwards. On the layer codes we tested, our schedules achieve a lower LER at roughly half LRC's CNOT depth per round. Consistent with this picture, the LER gap is present under a depth-agnostic noise model, where circuit depth carries no penalty, and several times larger under depth-aware noise (see \cref{fig:sec-ler-lrc}).

As an example, on the YBW26 $[[4,2,2]]$ code family with $\chi=(d,d,d)$, constructing the SEC for $d=4,6,8$ (148, 364, and 676 data qubits, respectively) takes 4.0\,s, 122\,s, and 179\,s to a valid circuit at proven-optimal depth, against 7.5\,s, 33\,s, and 149\,s for LRC's full construction, timed on the same machine in the same run. The schedule itself is the cheap part: the scheduling step reaches that circuit in 1.0\,s, 4.9\,s, and 24\,s, and the remainder is the minimum-weight ILP that supplies the logical representatives the rule aims at, which depends on the code rather than on the schedule. A further minute of the optional idle optimization removes $94\%$, $100\%$, and $91\%$ of all live ancilla idle at $d=4,6,8$, at least $90\%$ of what it reaches at any budget; its time budget is tunable, and spending the full default budget buys the last few percent of idle at a build time well above LRC's. On this family LRC's SEC round has a CNOT depth of 11 at every distance; ours is 5 at $d=4$ and 6 at $d=6$ and $d=8$.

The LER comparison of the same example is shown in \cref{fig:sec-ler-lrc}. At a fixed physical error rate, under the depth-aware per-tick model LRC's LER is $6$--$13$ times ours at $d=4$, $18$--$32$ at $d=6$ and $25$--$75$ at $d=8$ (X memory, statistically resolved points); the Z-basis comparison, run against LRC, shows the same pattern at slightly compressed margins. Under the depth-agnostic per-round model, which charges nothing for the difference in depth, the gap is a factor of a few: up to $6.5\times$ at $d=6$ and $11\times$ at $d=8$. Most of the advantage under the per-tick model therefore follows from the shallower round.

\begin{figure*}[t]
    \centering
    \includegraphics[width=\textwidth]{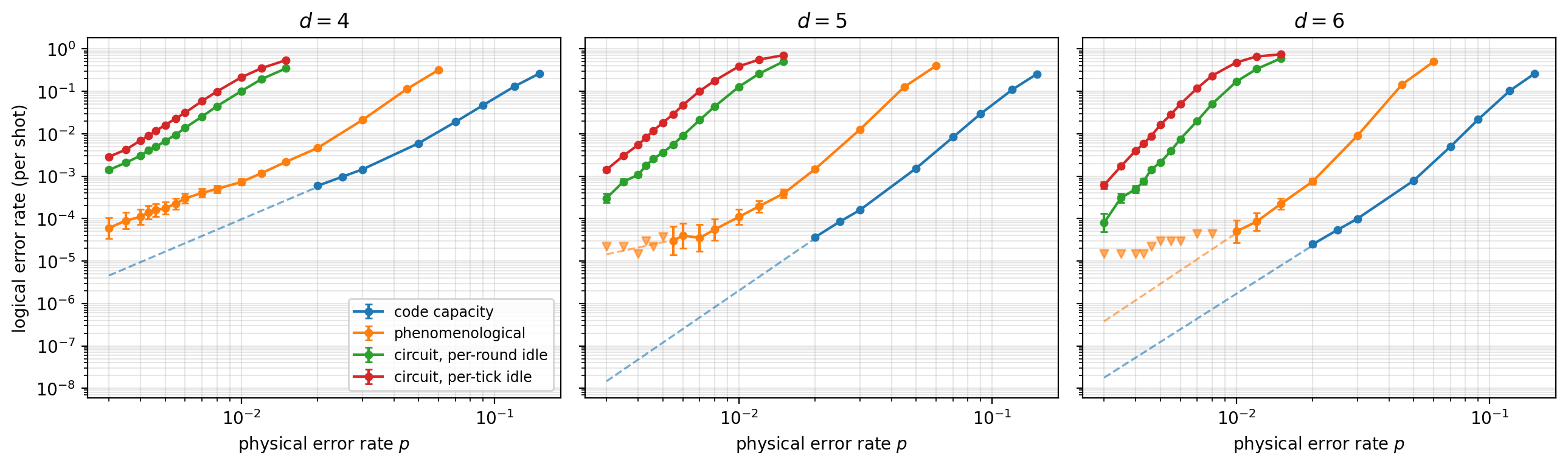}
    \caption{Cost of each successive noise layer for our SEC, on the YBW26 $[[4,2,2]]$ layer code family with $\chi=(d,d,d)$, $d=4,5,6$ (panels): X-basis memory, $d$ rounds, decoded with a global BP+OSD decoder on the raw detector error model (CSS-marginal detector convention). Each curve isolates one noise regime on a shared physical depolarizing parameter $p$: code capacity (data errors only), phenomenological (code capacity plus measurement errors), and circuit level (full circuit-level depolarizing noise) in two idle-noise variants: per-round (depth-agnostic) and per-tick (depth-aware, the default model). Successive layers shift the threshold down ($p_{\mathrm{th}}\!\approx\!0.15,\,0.040,\,7.3\times10^{-3},\,5.3\times10^{-3}$, respectively), so the gaps quantify the individual costs of measurement, gate/hook, and depth-dependent idle noise; the Z memory comparison (not shown) is quantitatively similar, with apparent thresholds within a few percent of the X memory values. Error bars are Wilson 95\% intervals. For points with fewer than five observed logical errors, we plot the one-sided 95\% upper confidence limit using a downward-pointing triangle. Dashed lines are sub-threshold power-law extrapolations of the code-capacity and phenomenological floors (a visual guide, not sampled data).}
    \label{fig:sec-benchmark}
\end{figure*}

We also benchmark the performance of our SEC by comparing it under several noise models: code capacity (data errors only), phenomenological noise (adding measurement errors), and full circuit-level noise, the latter under both a depth-agnostic (per-round) and a depth-aware (per-tick) idle model. \Cref{fig:sec-benchmark} shows the resulting comparison for the $[[4,2,2]]$ family at $d=4,5,6$. Every noise model exhibits a clean threshold, and the apparent thresholds, extracted from multi-distance crossings over the full $d=4$--$8$ range, are $p_{\mathrm{th}}\approx 0.15$, $4.0\times10^{-2}$, $7.3\times10^{-3}$, and $5.3\times10^{-3}$, respectively. Reading these thresholds as an error budget for the circuits we use, we find that the largest single cost is the gate and hook errors introduced by the CNOT circuitry, which lower the threshold by a factor of ${\sim}5.5$ relative to the phenomenological model; measurement errors cost a factor of ${\sim}3.7$ relative to code capacity, and the depth-dependent idle exposure a further ${\sim}1.4$. For these circuits gate and hook errors therefore dominate, and they are not addressable through depth; the residual idle term bounds what further depth work could buy, since idle cost scales directly with the SEC depth. These threshold ratios, however, understate the depth cost at realistic operating points: at a fixed near-threshold rate ($p\approx 4\times10^{-3}$, $d=6$), the LER of the depth-aware model is roughly $6$--$8$ times that of the depth-agnostic one, because this $p$ sits much nearer the per-tick threshold than the per-round one.

Beyond LER, we also evaluate our SEC through its circuit-level distance. The exhaustive search via QEDist~\cite{lee2026qedist} determines $d_{\mathrm{circ}}$ exactly for 35 of the 36 combinations of code and memory basis in \cref{tab:arms-layer}. In 32 cases, $d_{\mathrm{circ}}$ equals the corresponding code distance, indicating that the schedule incurs no distance loss due to hook errors. These include all codes with $[[4,2,2]]$ or $[[6,4,2]]$ inputs and all but one code in X memory. The three exceptions are the Steane-input codes at $\chi=(3,7,3)$, $(4,7,4)$, and $(5,4,5)$ in Z memory, for which $d_{\mathrm{circ}}$ is one less than $d_X$. For the remaining case, the Shor-input code at $\chi=(4,6,6)$ in X memory, we establish the bounds $11\le d_{\mathrm{circ}}\le d_Z=14$.

All experiments in the remainder of this paper use this SEC: the hook-constrained, depth-compressed schedule with mid-circuit detectors on memory-basis stabilizers, decoded globally on the raw detector error model, with noise conventions as stated per experiment.

\section{Experiments}
\label{sec:experiments}
\subsection{Memory thresholds}
\label{sec:memory-ler}

We first characterize the quantum-memory performance of several representatives under our SEC. We study four YBW26 layer code families, each a sequence of codes obtained by tuning $\chi$ at a fixed input code: the $[[4,2,2]]$ family at $\chi=(d,d,d)$ for $d=4,\dots,8$, for which $(d_X,d_Z)=(d,d)$; the $[[6,4,2]]$ family at $\chi=(d,6,d)$ for $d=4,\dots,8$, likewise with $(d_X,d_Z)=(d,d)$; a Steane-$[[7,1,3]]$ family of five codes, $\chi=(3,4,3),(3,7,3),(4,4,4),(4,7,4),(5,4,5)$, spanning $(d_X,d_Z)=(7,6)$ to $(11,10)$; and a Shor-$[[9,1,3]]$ series, $\chi=(2,6,4),(3,6,5),(4,6,6)$, with strongly asymmetric distances $(d_X,d_Z)=(4,8),(5,11),(6,14)$. All quoted distances are exact, certified by integer-programming computations on the codes' parity-check matrices~\cite{gurobi} rather than estimated from a decoder.

For every code we run X and Z memory experiments as defined in \cref{sec:background}. We use the circuit-level depolarizing noise model with per-tick (depth-aware) idle, and attach detectors following the CSS-marginal convention; the layer codes use our SEC circuits. We decode using global BP+LSD on the raw DEM; BP+OSD is our accuracy reference, but its per-shot cost grows steeply with circuit size, so we reserve it for cross-checks. We sample the LER as a function of the physical error rate $p$ across each family's crossing region and extract the asymptotic threshold by a finite-size scaling collapse: the curves of all codes in a family are fit jointly to a quadratic in the rescaled variable $x=(p-p_{\mathrm{th}})\,d^{1/\nu}$~\cite{wang2003confinement}, with $p_{\mathrm{th}}$ and $\nu$ as fit parameters. Here $d$ is the distance governing the basis in question ($d_Z$ for X memory, $d_X$ for Z memory); for the $[[4,2,2]]$ and $[[6,4,2]]$ families the two coincide. Simulation budgets, decoder settings and the parameters of every code are collected in \cref{app:methods}.

\begin{figure*}[t]
    \centering
    \includegraphics[width=0.92\textwidth]{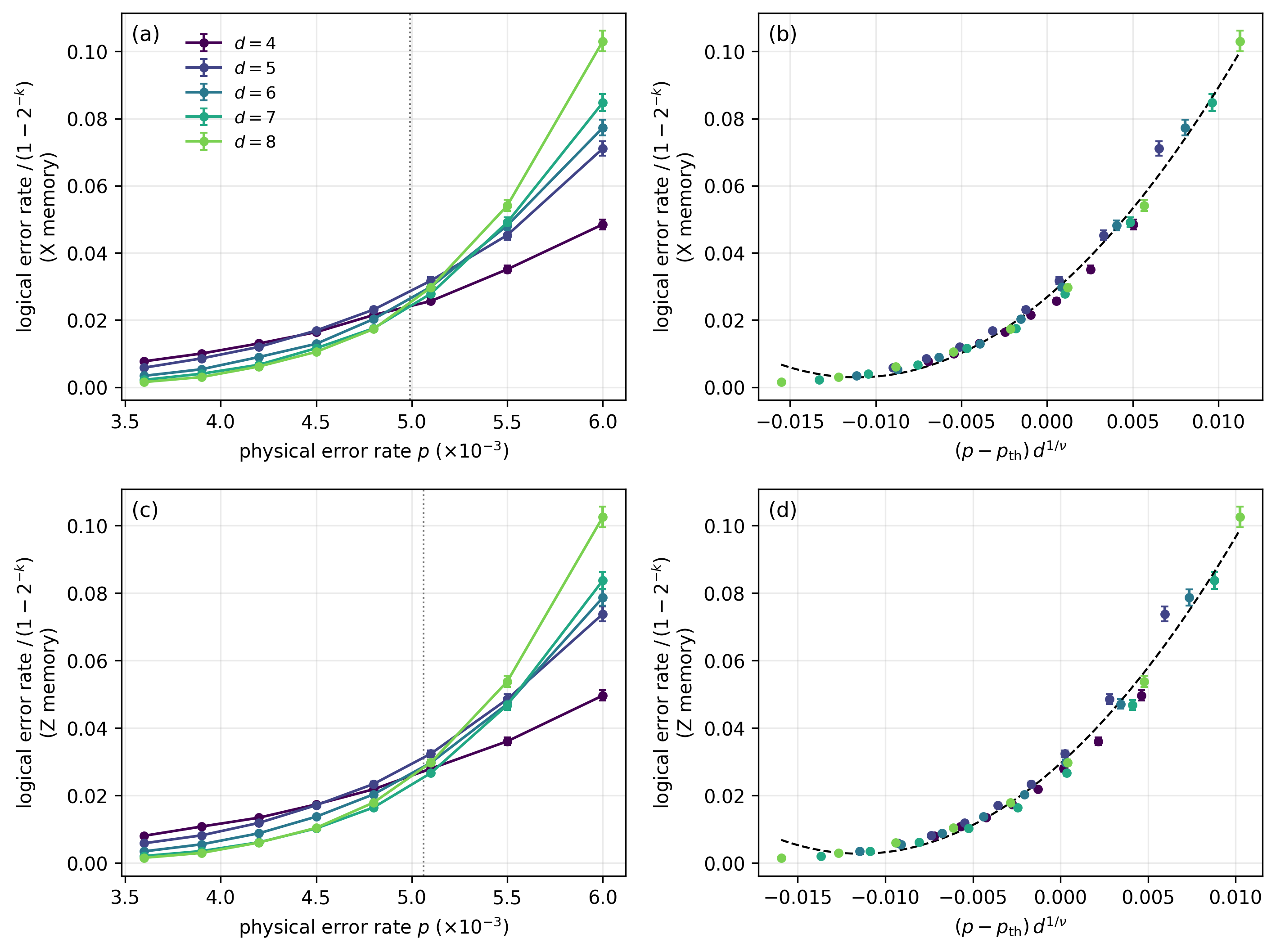}
    \caption{Memory threshold of the $[[4,2,2]]$ family, $\chi=(d,d,d)$, $d=4$--$8$ ($d_X=d_Z=d$): $d$ rounds of our SEC, circuit-level depolarizing noise with per-tick (depth-aware) idle, CSS-marginal detector convention, decoded with global BP+LSD on the raw DEM. Rates are normalized by the saturation value $1-2^{-k}$ throughout. (a),(c): logical error rate versus physical error rate $p$ across the crossing region for X and Z memory, respectively, one curve per distance $d$; error bars are Wilson 95\% intervals, and the dotted vertical line marks the fitted $p_{\mathrm{th}}$. (b),(d): the same data against the rescaled variable $(p-p_{\mathrm{th}})\,d^{1/\nu}$ with the joint quadratic fit (dashed), a local description of the crossing region that is not meaningful in the far tails, giving $p_{\mathrm{th}}=4.99(6)\times10^{-3}$ (X memory) and $5.06(7)\times10^{-3}$ (Z memory) with $\nu=0.86(8)$ and $0.87(9)$ respectively.}
    \label{fig:memory-threshold}
\end{figure*}

The $[[4,2,2]]$ family at $\chi=(d,d,d)$ is our primary threshold result (\cref{fig:memory-threshold}). This family is genuinely self-similar: every code is the same construction rescaled, and its five curves collapse cleanly onto a single scaling function with no outlier, giving X and Z memory thresholds of $p_{\mathrm{th}}=4.99(6)\times10^{-3}$ and $5.06(7)\times10^{-3}$, with $\nu=0.86(8)$ and $0.87(9)$ respectively.

Two further families test this value. The $[[6,4,2]]$ input code is closely related to $[[4,2,2]]$, but for its layer code family we set $\chi=(d,6,d)$ rather than the self-similar $\chi=(d,d,d)$. This scales only the two distance-carrying dimensions $\chi_X$ and $\chi_Z$; $\chi_Q$ contributes qubit overhead but no distance, and is held at the minimum value this input code admits, $\chi_Q=6$. This gives $p_{\mathrm{th}}=5.40(4)\times10^{-3}$ (X) and $5.51(6)\times10^{-3}$ (Z) with a clean collapse. The Steane family gives $p_{\mathrm{th}}=4.19(30)\times10^{-3}$ (X) and $4.10(35)\times10^{-3}$ (Z), with uncertainties several times larger, and two of its five codes sit off the collapse in both bases, $\chi=(4,4,4)$ above and $\chi=(4,7,4)$ below; they carry most of the quoted uncertainty: excluding $\chi=(4,4,4)$ tightens the fit to $4.30(18)\times10^{-3}$ (X) and $4.23(21)\times10^{-3}$ (Z), while excluding $\chi=(4,7,4)$ moves it the other way, to $3.92(25)\times10^{-3}$ and $3.80(28)\times10^{-3}$. The $[[4,2,2]]$ and $[[6,4,2]]$ families agree to within $9\%$, at $5.0$--$5.5\times10^{-3}$, so $\chi$-tuning yields a convergent circuit-level threshold across both. The Steane central values sit $16$--$26\%$ lower, two to three times that family's own fit uncertainty.

The Shor family behaves differently. The $[[9,1,3]]$ input code is strongly X/Z-asymmetric (X-checks have weight 6 and Z-checks have weight 2), and this asymmetry propagates through the construction to the threshold itself: in X memory ($d_Z=8,11,14$) the series collapses to $p_{\mathrm{th}}=3.13(14)\times10^{-3}$, whereas in Z memory ($d_X=4,5,6$) it gives $5.08(29)\times10^{-3}$, a fitted compromise across this non-self-similar series whose three pairwise crossings straddle the sampled range, from below $3.7\times10^{-3}$ to above $6.4\times10^{-3}$.

As a reference we ran two types of surface code through the identical pipeline: same noise model, detector convention, decoder and collapse fit. The rotated one is Stim's generated circuit with its basis-conjugating Hadamards absorbed into native-basis resets and measurements, matching the ancilla conventions of every other circuit; the unrotated one comes from our own builder with a balanced alternating CX schedule~\cite{orourke2025comparepair}. The unrotated surface code gives $p_{\mathrm{th}}=8.02(6)\times10^{-3}$ (X) and $8.29(6)\times10^{-3}$ (Z), the rotated one $7.71(9)\times10^{-3}$ and $7.73(14)\times10^{-3}$; all uncertainties quoted here are fit covariance only. Two control experiments in the same setting support these values: refitting both families under BP+OSD moves the fitted thresholds by less than $0.6\%$ on the unrotated surface code and by $1$--$2\%$ upward on the rotated one, the direction expected from BP+LSD's slight pessimism (\cref{app:lsd-osd}), so the choice of decoder does not affect the comparison; and swapping the unrotated patches' balanced schedule for a single fixed one leaves the basis-averaged threshold unchanged to $0.2\%$ while swinging the X/Z split by ${\sim}8$ percentage points, so per-basis splits on the surface codes are schedule artifacts rather than code properties.

Averaging over the two bases, at these distances the $[[4,2,2]]$ family therefore sits a factor of $1.54$ below the rotated surface code and $1.62$ below the unrotated one. The unrotated comparison is the informative one: unrotated patches are the layer code's own constituent patches, so quasi-concatenation retains roughly three fifths of the threshold of its building block.

The comparison above charges idle exposure at every circuit tick, so it prices the depth of each family's syndrome extraction. A depth-agnostic comparison isolates that contribution: applying the idle channel once per round instead of once per tick, the cross-family gap nearly closes, with the $[[4,2,2]]$ family at $p_{\mathrm{th}}=7.01(8)/7.02(9)\times10^{-3}$ (X/Z) against $6.98(12)/7.18(10)$ for the unrotated surface code and $6.89(15)/6.81(21)\times10^{-3}$ for the rotated one, all within a few percent of each other. The per-tick threshold gap therefore comes mainly from the layer code's deeper syndrome-extraction round, not from the quality of its hook-error structure. Owing to a finite-size effect, the surface codes' thresholds are lower under this model although their LER is lower at every sampled point. A surface code's boundary qubits take part in fewer CNOTs, so per-tick idle noise charges them more than the bulk ones. The boundary is a larger share of a smaller patch, so this raises the LER of small codes more than that of large ones and moves the crossing of their curves to higher $p$. This effect shrinks as codes with larger $d$ enter the fit.

Three caveats qualify the reported numbers. First, on layer code circuits BP+LSD is measured to be slightly pessimistic in the error-rich fitting regime (quantified in \cref{app:lsd-osd}); this can only bias the fitted thresholds downward, so the values above are conservative. A same-seed BP+OSD run alongside BP+LSD on the $[[4,2,2]]$ noise-model benchmark of \cref{sec:sec-design} measures the bias directly at the same $p$: over all resolved points the ratio has median $1.17$ at $d=4$, rising to $1.36$ at $d=8$, always with BP+LSD on the high side. Second, the quoted uncertainties are the statistical errors of the collapse fit alone. Excluding a single code moves $p_{\mathrm{th}}$ by more than that on every family except the Steane family, where the two are comparable, so the total uncertainty is generally set by the fit-choice spread rather than by the statistical errors. Third, the exponents $\nu$ are descriptive parameters of a fit over a short distance range ($4$--$8$ for the $[[4,2,2]]$ and $[[6,4,2]]$ families, $6$--$11$ across the two bases for the Steane family and $4$--$14$ for the Shor series); we attach no universality claim to them.

\subsection{Idle robustness}
\label{sec:idle}

In fault-tolerant surface-code architectures, logical qubits are protected by repeated rounds of syndrome extraction and decoding, which continuously suppress logical errors. When used as a quantum memory, the yoked surface code \cite{gidney2025yoked} introduces an outer concatenated ``yoke'' code whose parity checks are measured only periodically. The interval between successive yoke measurements can therefore be viewed as an effective idle period for the outer code, while the inner surface code syndrome extraction continues uninterrupted. This naturally motivates asking a similar question for the quasi-concatenated layer code construction: how infrequently can syndrome-extraction rounds be scheduled while maintaining a fixed LER, and does a layer code permit longer idle intervals than a surface code of the same distance under otherwise identical conditions? To our knowledge, no direct comparison of this form has been reported. Existing circuit-level memory benchmarks \cite{googleqec2024below} typically vary the number of \emph{active} syndrome-extraction rounds rather than introducing SEC-free idle intervals between them.

We measure this with the sustained-memory circuit (\cref{eq:sustained}): after preparing an X-basis memory, each of $N$ refresh cycles applies $r$ rounds of our SEC and then holds the data idle for $t$ ticks of single-qubit depolarizing noise at the same physical rate $p$, and a final transversal readout is decoded once, globally, with BP+LSD on the full spacetime raw DEM, under the same per-tick idle and CSS-marginal detector conventions as \cref{sec:memory-ler}. Throughout, a refresh cycle means this SEC-plus-idle period, not a single syndrome-extraction round.

\begin{equation}
\fiteqn{
\underbrace{\text{prep}}_{d_{Z}\ \text{rounds}}
\to \Bigl[\,
   \underbrace{\text{SEC}}_{r\ \text{rounds}}
   \to
   \underbrace{\text{idle}}_{t\ \text{ticks},\,\mathrm{DEP1}(p)}
\,\Bigr]^{N}
\to \underbrace{\text{readout}}_{\text{transversal}}
\to \text{decode}
}
\label{eq:sustained}
\end{equation}

During each idle window the accumulated noise composes into a net single-qubit depolarizing channel, so the readout-relevant accumulated flip probability $q_{\mathrm{eff}}(p,t)$ is known in closed form:
\begin{equation}
q_{\mathrm{eff}}(p,t) = \tfrac{1}{2}\Bigl(1-\bigl(1-\tfrac{4p}{3}\bigr)^{t}\Bigr)
\label{eq:qeff}
\end{equation}

It is the probability that the accumulated Pauli on one data qubit anticommutes with the readout basis, and depends only on $p$ and $t$, not on the code.

This idle channel is energy-blind, making no reference to a code Hamiltonian and thus admitting no energy barrier. This experiment therefore measures robustness to accumulated noise under the standard circuit-level noise model, rather than the thermal partial self-correction considered in prior work.

Assume each refresh cycle has probability $\lambda(r,t)$ of randomizing the stored logical state. Separating out the $N$-independent contribution of preparation and readout, the $k$-normalized LER of the process in \cref{eq:sustained} can then be described by
\begin{equation}
    \frac{\mathrm{LER}(N,r,t)}{1-2^{-k}}\approx 1-(1-\lambda(r,t))^{N}
    \label{eq:percycle}
\end{equation}

An alternative model instead lets each of the $k$ logical observables flip independently with probability $f$ per cycle~\cite{googleqec2024below}. The two models coincide for $k=1$, with $\lambda=2f$, and agree to first order for $k=2$, so we restrict the fit to the small-error regime $\mathrm{LER}/(1-2^{-k})\le 0.3$. We verified that switching between the two models changes every reported interval by less than 1\%.

Our primary quantity is the sustainable SEC-free idle interval $t^*$, which is defined as the idle time at which $\lambda$ reaches a budget of $0.01$. The fit assumes the refresh cycles are independent and alike, which we verify below. The ratios of $t^{*}$ across families are the intervention frequency comparison. At a fixed physical error rate $p$, comparing codes at the same idle duration $t$ is equivalent to comparing them at the same accumulated flip probability $q_{\mathrm{eff}}$, so these ratios are independent of the physical duration assigned to a tick. A longer interval does not imply less control work: at the same distance the layer code runs $1.7$--$5.0$ times more two-qubit gates, measurements and noisy locations per logical qubit per unit time, a ratio that grows with distance. The $t^{*}$ comparison is therefore about how often correction is applied, not what it costs.

The detailed method is as follows. Fixing $p=10^{-3}$, roughly a fifth of the layer code thresholds of \cref{sec:memory-ler}, and using X memory, we first collect LER for each code and then fit the curve $\lambda(t)$ for a fixed $r$ according to \cref{eq:percycle}. From $\lambda(t)$, the sustainable idle interval $t^{*}$ can then be extracted. We sweep the refresh depth $r\in\{1,2,3\}$, and the idle time $t$ along a geometric grid targeting a common $q_{\mathrm{eff}}(p,t)$ band (up to $q_{\mathrm{eff}}\approx0.04$, extended for the largest codes), which is chosen to bracket the $\lambda=0.01$ crossing; at each $(r,t)$ we further sweep the refresh-cycle count $N\in\{2,4,8,16\}$. We also probe the case $N=1$, $r=0$ and plot the LER against $q_{\mathrm{eff}}(p,t)$ as a favorable reference. At $r=1$ we further trace the low-$p$ trend, sampling $p\in\{2.5,5\}\times10^{-4}$ and $2.5\times10^{-3}$, and repeat the comparison in Z memory.

The layer codes are drawn from the families of \cref{sec:memory-ler}: the $\chi=(d,d,d)$ codes of the $[[4,2,2]]$ family at $d=4,6,8$ ($k=2$) and that section's Steane-input code $\chi=(3,4,3)$ ($(d_X,d_Z)=(7,6)$, $k=1$). They are compared with the rotated and unrotated surface codes of \cref{sec:memory-ler} at $d=4,6,8$, adapted to identical conventions and setup. We report the numbers under the any-of-$k$ LER convention defined in \cref{sec:decoding}. Re-analyzing the same data under a per-logical-worst convention shifts the $k=2$ layer codes' intervals by $11$--$15\%$ at $d=4$ and by at most $5\%$ at $d=6,8$.

\begin{figure*}[t]
    \centering
    \includegraphics[width=0.98\textwidth]{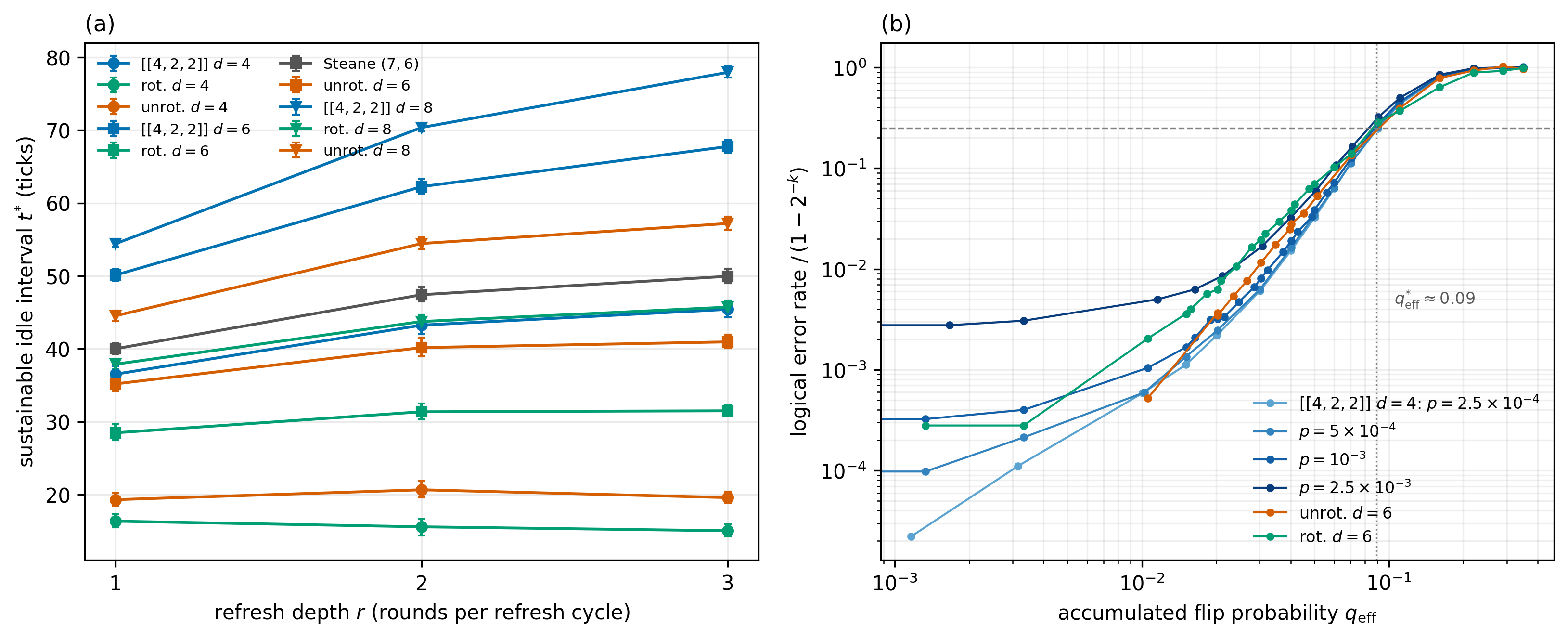}
    \caption{Idle robustness of the sustained memory. (a) Sustainable idle interval $t^{*}$ (budget $\lambda=0.01$ per refresh cycle) versus refresh depth $r$, at $p=10^{-3}$, X-basis memory: colour denotes code family and marker shape the distance, and error bars propagate the $95\%$ interval on $\lambda(t)$ through the crossing. The full ordering (rotated $<$ unrotated $<$ layer at equal distance) holds at every $r$, and for the layer codes $t^{*}(r)$ increases with $r$ over $r=1,2,3$. The single-window benchmark of \cref{tab:idle} is a separate experiment and is not shown here. (b) The $q_{\mathrm{eff}}$ collapse on the single-window experiment, which establishes the accumulated-noise axis (logical error rate normalized by its large-noise saturation value $1-2^{-k}$; dashed line at $0.25$): the four $[[4,2,2]]$ $d=4$ curves at different $p$ merge onto one function of $q_{\mathrm{eff}}$ once idle noise dominates and fan into $p$-dependent floors below it, and the different codes cross the loose diagnostic level of $0.25$ used here, well above the $\lambda=0.01$ operating budget of \cref{tab:idle}, at a nearly code-independent $q^{*}_{\mathrm{eff}}\approx0.09$ (dotted line). Points with fewer than five observed logical errors are omitted.}
    \label{fig:idle}
\end{figure*}

\Cref{tab:idle} gives the sustainable intervals $t^{*}$ at $p=10^{-3}$, and \cref{fig:idle}(a) plots them against refresh depth $r$. At equal distance the ordering rotated $<$ unrotated $<$ layer holds at every $r$ we tested, and $t^{*}$ increases with $d$ within each family.

We can now check the independence assumption stated above. At small $r$ one refresh cycle can leave residual error behind for the next, so we compare the $\lambda$ implied by the $N=2$ point with the one implied by $N=16$, at the sampled idle time nearest $t^{*}$: if the refresh cycles are alike the two values should agree, and we call the fit stationary. On the $d=8$ layer code they differ by a factor $5.1$ at $r=1$, but only $1.8$ at $r=2$ and $1.3$ at $r=3$, and agreement sets in at lower $r$ for smaller codes. Taking agreement to mean a factor within $1.5$, the minimum useful refresh depth is $r\approx2$, $2$, and $3$ at $d=4$, $6$, and $8$. All are far below the distance itself, so a few SEC rounds per refresh cycle already give the global decoder enough time-like redundancy to sustain the interval.

Of all the codes, the $d=8$ layer code is the last to become stationary, doing so only at $r=3$, the largest depth we simulate. We therefore quote the intervals there. The layer codes' sustainable intervals are then $3.0\times$, $2.2\times$, and $1.7\times$ longer than the rotated surface code at $d=4$, $6$, and $8$, and $2.3\times$, $1.7\times$, and $1.4\times$ longer than the stronger unrotated surface code. The advantage thus holds at every distance but narrows as $d$ grows.

\begin{table}[t]
\begin{ruledtabular}
\begin{tabular}{lccccc|c}
code & $n$ & $(d_X,d_Z)$ & $r{=}1$ & $r{=}2$ & $r{=}3$ & \makecell{single\\window} \\
\hline
rotated $d{=}4$ & 31 & $(4,4)$ & 16.3 & 15.6 & 15.0 & 15.9 \\
unrotated $d{=}4$ & 49 & $(4,4)$ & 19.3 & 20.7 & 19.6 & 19.1 \\
$[[4,2,2]]$ $d{=}4$ & 294 & $(4,4)$ & 36.5 & 43.2 & 45.4 & 50.4 \\
rotated $d{=}6$ & 71 & $(6,6)$ & 28.5 & 31.4 & 31.5 & 35.9 \\
unrotated $d{=}6$ & 121 & $(6,6)$ & 35.2 & 40.2 & 40.9 & 37.2 \\
Steane $\chi{=}(3,4,3)$ & 385 & $(7,6)$ & 40.0 & 47.4 & 49.9 & 55.7 \\
$[[4,2,2]]$ $d{=}6$ & 726 & $(6,6)$ & 50.1 & 62.3 & 67.8 & 74.1 \\
rotated $d{=}8$ & 127 & $(8,8)$ & 37.9 & 43.7 & 45.7 & 47.9 \\
unrotated $d{=}8$ & 225 & $(8,8)$ & 44.5 & 54.4 & 57.2 & 59.4 \\
$[[4,2,2]]$ $d{=}8$ & 1350 & $(8,8)$ & 54.4 & 70.4 & 77.9 & 83.9 \\
\end{tabular}
\end{ruledtabular}
\caption{Sustainable idle interval $t^{*}$ (in ticks; log-interpolated crossing of $\lambda(t)$) at the budget $\lambda=0.01$ per refresh cycle, for X-basis memory at $p=10^{-3}$, as a function of refresh depth $r$. $n$ is the total physical qubit count of the syndrome-extraction circuit, data plus one ancilla per stabilizer; X memory is governed by $d_Z$, and all codes except the Steane-input one have $d_X=d_Z$. The final column is the single-window benchmark ($N=1$, $r=0$), in which the idle interval ends directly in transversal data readout: a favorable perfect-terminal-syndrome boundary, shown for reference only, and not a bound (in the unrotated $d=6$ row, the $r=2$ and $r=3$ intervals slightly exceed it). For the layer codes the sustained intervals rise toward it with $r$ over the tested $r=1,2,3$ ($[[4,2,2]]$ $d=8$: $65\%$, $84\%$, and $93\%$ of it; $d=6$: $68\%$, $84\%$, and $91\%$). Among the surface codes the response to $r$ is weaker: most gain little beyond $r=2$, and the smallest rotated one declines with $r$.}
\label{tab:idle}
\end{table}

The single-window benchmark ($N=1$, $r=0$) reduces to a single idle window followed directly by the transversal readout, so the accumulated idle noise is decoded against a perfect terminal syndrome and neither refresh rounds nor preceding refresh cycles contribute faults. It is therefore a favorable reference rather than a mode of operation: a stored logical qubit must be held across many refresh cycles, each paying for its own syndrome extraction. The operationally meaningful values remain the sustained intervals of \cref{tab:idle}.

The result is robust in three further directions. First, the sustained advantage widens as $p$ decreases. At $r=1$, as $p$ falls from $10^{-3}$ to $5\times10^{-4}$, the $[[4,2,2]]$ $d=6$ interval grows from $1.42\times$ to $1.51\times$ that of unrotated $d=6$, and from $1.76\times$ to $1.85\times$ that of rotated $d=6$; at $2.5\times10^{-4}$ the layer code stays below budget over the whole sampled grid, leaving only the lower bounds $1.45\times$ and $1.82\times$. This is expected rather than a new effect: when $p$ is small, $q_{\mathrm{eff}}$ in \cref{eq:qeff} reduces to $\tfrac{2}{3}pt$, so at a fixed tolerated $q^{*}_{\mathrm{eff}}$ the crossing scales as $t^{*}\!\propto\!q^{*}_{\mathrm{eff}}/p$ and the ratio of intervals tends to the ratio of tolerated $q^{*}_{\mathrm{eff}}$. Second, the ordering is unchanged in Z memory at $r=1$. On the three $d=6$ codes, where $d_X=d_Z$, $t^{*}$ stays within $5\%$ of its X memory value; on the Steane-input code it is $7\%$ longer. Third, a BP+OSD cross-check on the $d=6$ layer code, with both decoders fitted on the same set of refresh-cycle counts, shifts its sustained interval by ${\sim}2\%$, in the direction that understates the layer advantage (\cref{app:lsd-osd}).

The $q_{\mathrm{eff}}$ collapse (\cref{fig:idle}(b)) confirms the mechanism: $p$ and $t$ enter only through the composed idle channel, so $t^{*}$ is simply the time at which $q_{\mathrm{eff}}$ reaches the accumulated-noise level a code tolerates. Those levels are what separate the families, e.g., at $d=6$ and $r=3$ the layer code withstands $q^{*}_{\mathrm{eff}}=0.043$, against $0.027$ for unrotated and $0.021$ for rotated. Thermal self-correction is ruled out by the energy-blind construction of the noise rather than by this collapse. The layer codes' advantage is thus a statement about the strength of their error-correction response to accumulated noise: the same quasi-concatenated geometry that produces a growing energy barrier, and hence partial self-correction, in the finite-temperature Hamiltonian setting~\cite{layercodepy2510,williamson2025partialselfcorrectionlayercodes} also produces the circuit-level idle robustness measured here; the two settings are complementary, and our results neither confirm nor probe the thermal memory lifetime.

Three caveats bound the scope of this advantage: the decoding model it assumes, the per-state normalization, and the even distances it compares.

First, the results above assume one offline global decode of the whole run. A real-time decoder instead commits in a forward sliding window of $W$ refresh cycles, and under it the layer advantage erodes: the layer code's low global error rate leans on time-like redundancy across refresh cycles, which a short window cannot pool. To quantify the erosion we re-decode the sustained runs with such a decoder and extract $t^{*}_{\mathrm{win}}$, the sustainable interval under the same $\lambda=0.01$ budget, quoted as a fraction of its offline-global value (\cref{tab:idle-window}). At each step such a decoder decodes the $W$ most recent refresh cycles, finalizes the correction on the oldest $c$ of them, and advances by $c$; we call $c$ the commit width, and test $c=1,2$.

The commit width matters as much as the window length. At $d=6$ and $r=2$ the layer code keeps $0.18$--$0.19$ of its global interval across $W=2$--$4$ when $c=1$, and $0.30$ when $c=2$; the surface codes keep $0.46$--$0.66$ (unrotated) and $0.69$--$0.84$ (rotated) under the same decoders, which leaves the layer code with the shortest windowed interval of the three. The penalty is distance-robust rather than distance-growing: at $d=8$ the two commit widths give $0.18$ and $0.29$. The erosion is a property of this protocol and this decoder: the two commit widths already differ by a factor of ${\sim}1.6$ on the layer code, so a windowed decoder with overlapping recovery or a longer commit could shrink the penalty further.

Second, these ratios compare one logical state with another at equal distance; they say nothing about qubit efficiency. The layer codes buy their longer intervals with roughly an order of magnitude more physical qubits than the rotated surface codes.

Third, every code compared here has even distance, so like is compared with like and the comparison stands as measured. It cannot, however, be turned into a fair footprint comparison: a distance-$d$ code fails only once $\lceil d/2\rceil$ faults occur, so an even-$d$ code is no better protected than its odd-$d$ predecessor while using more qubits. Such a comparison would have to run along each family's odd-distance frontier, which we do not attempt here.

\begin{table}[t]
\begin{ruledtabular}
\begin{tabular}{llccc}
code ($d=6$, $r=2$) & commit & $W{=}2$ & $W{=}3$ & $W{=}4$ \\
\hline
$[[4,2,2]]$ (layer) & $c{=}1$ & 0.19 & 0.19 & 0.18 \\
 & $c{=}2$ & $<0.08$ & 0.30 & 0.29 \\
unrotated & $c{=}1$ & 0.46 & 0.46 & 0.46 \\
 & $c{=}2$ & 0.45 & 0.66 & 0.65 \\
rotated & $c{=}1$ & 0.69 & 0.69 & 0.70 \\
 & $c{=}2$ & 0.64 & 0.84 & 0.83 \\
\end{tabular}
\end{ruledtabular}
\caption{Windowed-decoding retention $t^{*}_{\mathrm{win}}/t^{*}$: the sustainable idle interval of a forward sliding-window decoder of window length $W$ refresh cycles, as a fraction of the offline-global value of \cref{tab:idle}, at $d=6$, $r=2$, $p=10^{-3}$, X memory, for commit widths $c=1,2$. Cells marked $<0.08$ are censored: the windowed crossing falls below the sampled idle grid, so the entry is a one-sided bound. For $W{=}4$ the fit rests on the $N=8,16$ runs only: an $N\le W$ run is no longer than the window itself, so its decode is effectively global and carries no windowed information. The column is shown only as a consistency check on $W{=}3$.}
\label{tab:idle-window}
\end{table}

Overall, in a memory-equipped architecture where idle logical qubits are periodically refreshed by SEC, a layer code memory tolerates a longer idle interval between refresh rounds than a surface code of the same distance at a fixed failure budget per refresh cycle, and correspondingly needs fewer correction rounds per unit storage time. Against the rotated surface code the margin in $t^{*}$ is $1.7$--$3.0\times$ at $r=3$ over the $d=4$--$8$ range we simulate. We state the scope of this advantage rather than extrapolate it: it narrows with distance over the simulated range and reverses under the windowed decoders we test, so whether it persists to fault-tolerant distances is untested.

\section{Hardware implementation}
\label{sec:hardware}

\begin{figure*}[t]
    \centering
    \includegraphics[width=0.92\textwidth]{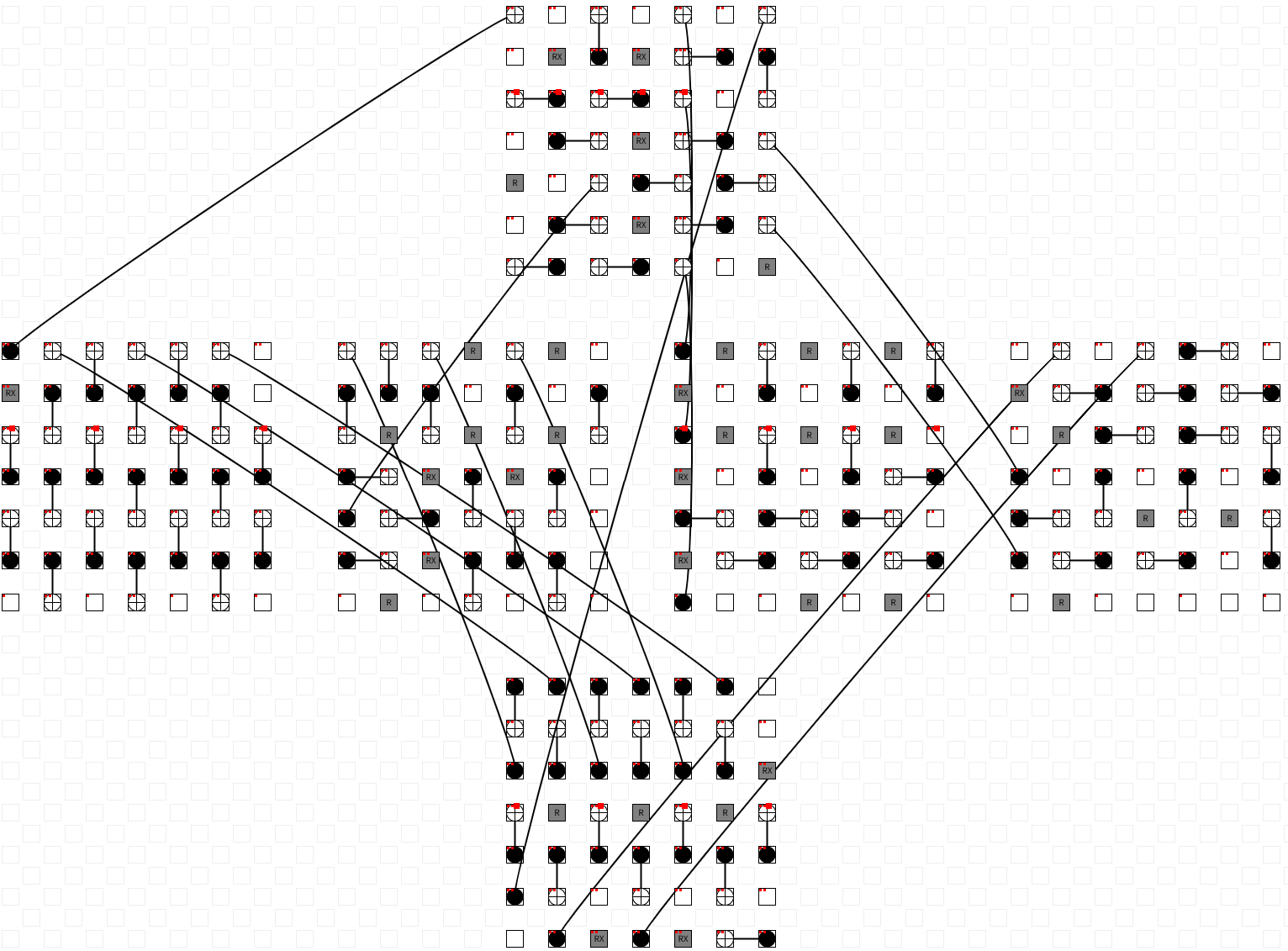}
    \caption{A single tick of the syndrome extraction circuit of the $d=4$ $[[4,2,2]]$ layer code at $\chi=(4,4,4)$, X-basis memory, drawn in Crumble~\cite{crumble}. The six surface code patches are arranged as in the right panel of \cref{fig:layer-codes}: one patch per input qubit in the middle row, and one patch per input check above and below. Small squares are physical qubits, and each segment is a CNOT executed in this tick, its filled end marking the control. The gates within a patch are the local ones of an ordinary unrotated surface code. The long segments running between patches are the junction CNOTs, which act between qubits belonging to different planes of the code and require long-range shuttling when implemented directly in this layout. Boxes labelled R and RX are ancilla resets, in the $Z$ and $X$ bases respectively.}
    \label{fig:tick}
\end{figure*}

If a layer code is implemented directly on a two-dimensional device, the junction CNOTs coupling different layer-code planes appear as long-range interactions. Such interactions can be realized by qubit shuttling on platforms such as trapped ions~\cite{RevModPhys.75.281,blatt2008entangled,kaushal2020shuttling,PhysRevX.12.011032}, neutral atoms~\cite{bluvstein2022quantum,evered2023high,bluvstein2024logical} and silicon spin qubits~\cite{loss1998quantum,buonacorsi2019network,PhysRevApplied.18.024053,PhysRevApplied.18.044064,de2025high}. \Cref{fig:tick} shows one tick of the syndrome extraction circuit of the $d=4$ $[[4,2,2]]$ layer code laid out directly in two dimensions.

One direct shuttling-based implementation of a cross-plane CNOT is to bring the participating qubits into a common interaction region, apply the gate, and return them afterwards. In such a direct implementation, the shuttling distance associated with a junction CNOT can grow with the separation of the coupled patches and hence with the code distance. The junction CNOT can therefore be substantially more expensive than an intra-patch CNOT in both latency and fidelity. The dominance of long-range shuttling cost can be seen in, for example, reconfigurable neutral-atom experiments where the characteristic free-space movement time between gates can be much longer than the entangling-gate time~\cite{bluvstein2024logical}. In a shuttling-based trapped-ion processor we can also see that shuttling accounts for $95\%$ of the execution time of a demonstrated fault-tolerant parity-check sequence~\cite{PhysRevX.12.011032}.

\begin{figure}[t]
    \centering
    \makebox[\linewidth][l]{\textbf{(a)}}
    \par\vspace{0.2em}
    \includegraphics[width=0.97\linewidth]{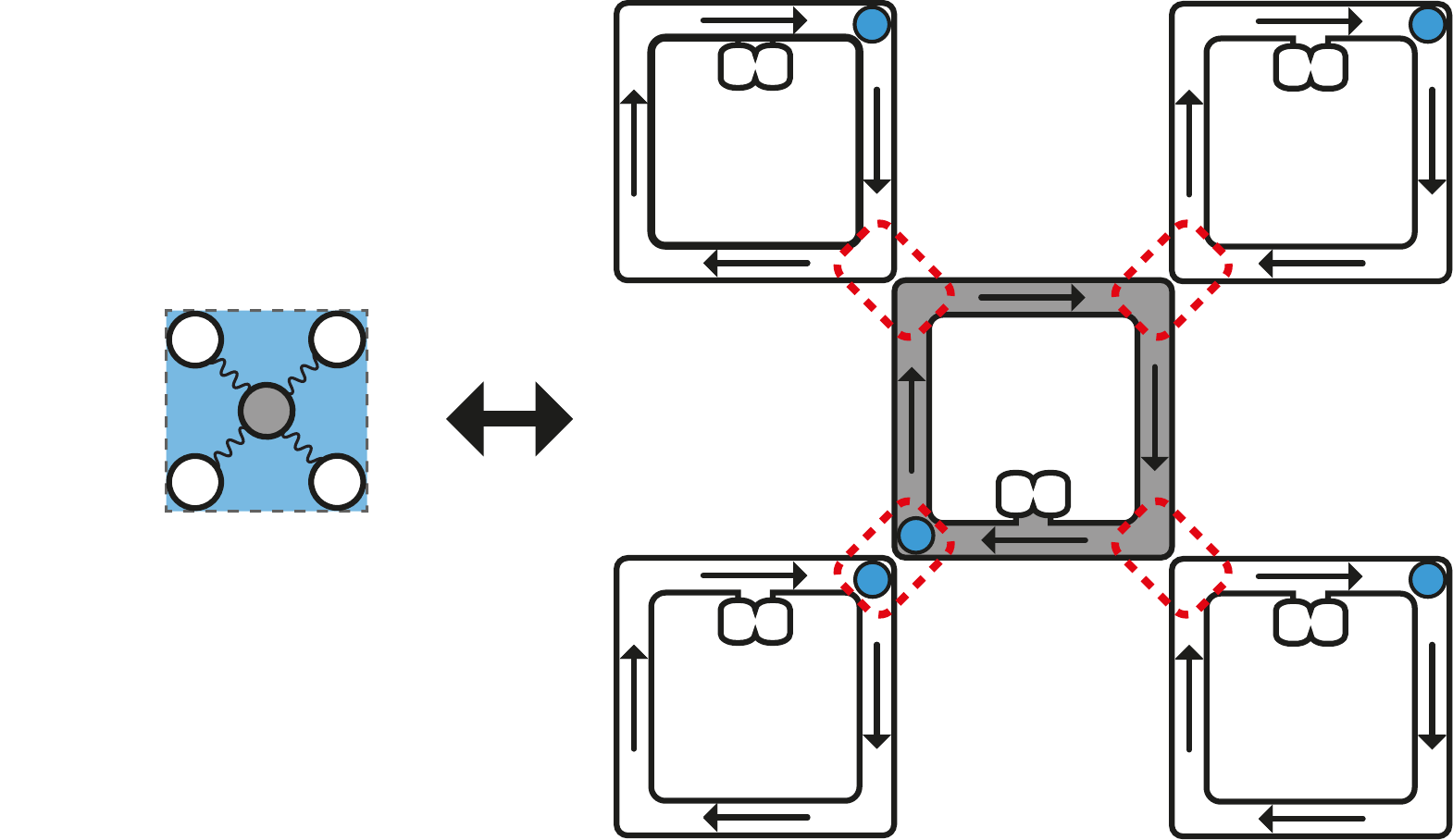}
    \par\vspace{1.0em}
    \makebox[\linewidth][l]{\textbf{(b)}}
    \par\vspace{0.2em}
    \includegraphics[width=\linewidth]{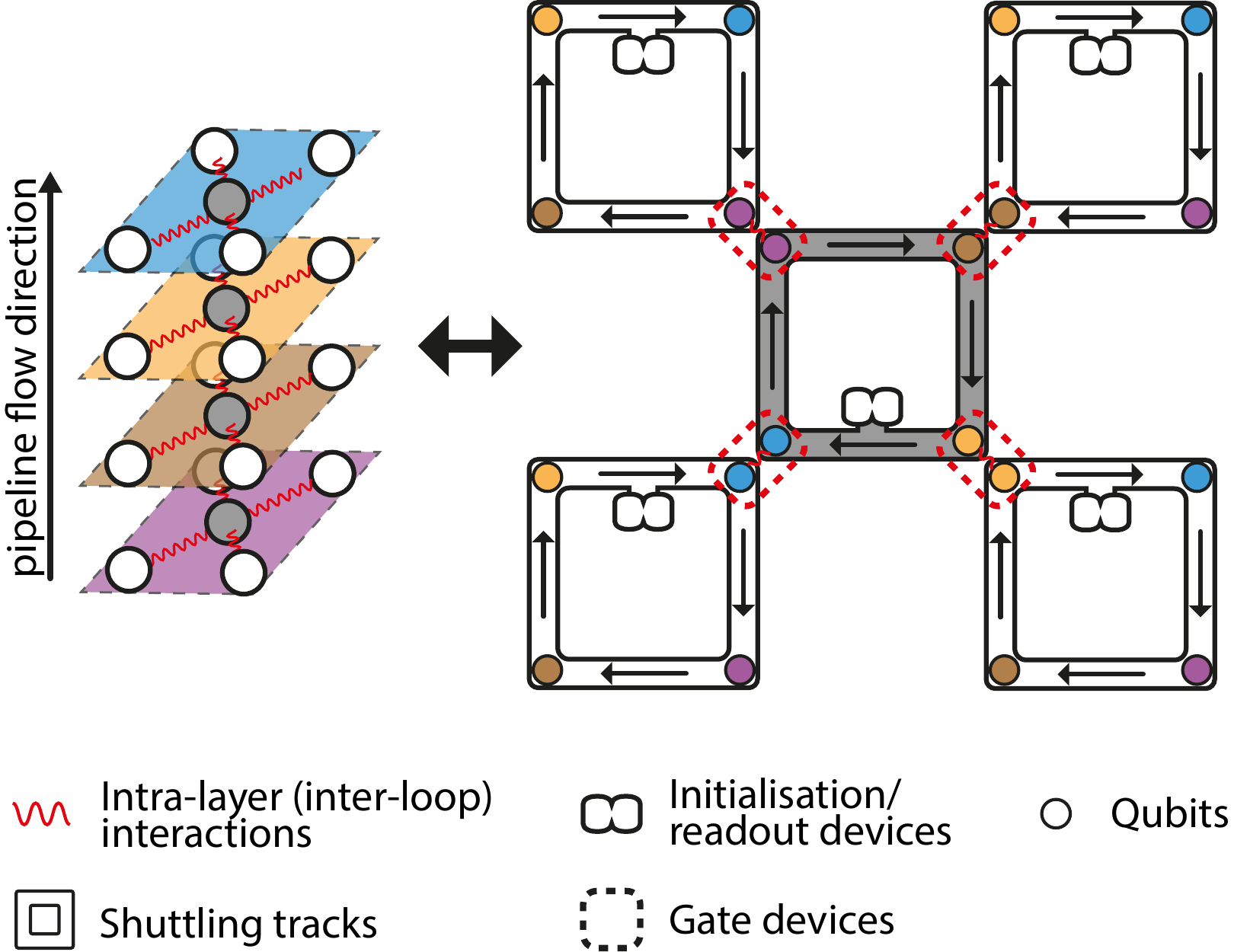}
    \caption{The looped pipeline architecture. (a) A single 2D qubit layer. The inset on the left is the abstract unit cell: a central qubit (grey) coupled to its four neighbours (white), with the wavy lines marking intra-layer, that is inter-loop, couplings. On the right is the hardware realization, in which every qubit is shuttled around a closed track (rounded squares) in the direction of the black arrows; the grey track carries the central qubit of the cell, and the blue dots mark the qubits themselves. Where two tracks approach each other, the red dashed boxes are gate devices that mediate the inter-loop coupling, and the paired-box symbols are initialization and readout devices. (b) The same set of tracks hosting a stack of four layers. Loading several qubits into each loop and shuttling them in a pipeline lets one loop carry one qubit per layer, coloured blue, yellow, brown and purple to match the four layers of the left inset and ordered along the pipeline flow direction. The number of stacked layers is then set by the pipeline occupancy rather than by the device layout, so no additional hardware is needed. Panels adapted from Fig.~4 of Ref.~\cite{PRXQuantum.4.020345}, published under a CC BY 4.0 license.}
    \label{fig:looped-pipeline}
\end{figure}

It is worth revisiting what connectivity the code actually demands and whether long-range shuttling is truly unavoidable. A layer code can, to a good approximation, be viewed as a collection of unrotated surface code patches coupled in a prescribed pattern. Within each patch the required connectivity is local; the additional requirement is the set of couplings between different patches or planes. A looped-pipeline architecture provides these couplings on any platform that supports qubit shuttling, using only short-range transport~\cite{PRXQuantum.4.020345,sun2026foldedsurfacecodearchitecture}. By shuttling the qubits around loops rather than along linear tracks and operating them as a pipeline, it realizes an effective stack of 2D qubit arrays and hence scalable 3D connectivity on a strictly 2D hardware platform, as illustrated in \cref{fig:looped-pipeline}. The planes of a layer code can therefore be stacked in a looped pipeline. Taking silicon spin qubits as an example, a shuttling loop is usually of $\mu$m scale, and single-electron spin shuttling has been demonstrated over an effective distance of $10\,\mu\mathrm{m}$ in under $200\,\mathrm{ns}$, preserving the spin with an average fidelity of $99.5\%$~\cite{de2025high}. This shuttling time is of a similar order to the state-of-the-art two-qubit gate times of around $100\,\mathrm{ns}$~\cite{mills2022two,xue2022quantum}. A carefully designed shuttling scheme can therefore implement intra-loop (cross-plane) and inter-loop (in-plane) CNOTs with roughly the same runtime~\cite{sun2026foldedsurfacecodearchitecture}.

However, in the absence of such a scheme, a junction CNOT is slower or noisier than a local one, and we now model that extra cost. We introduce two parameters: $\alpha=p_{\mathrm{junction}}/p_{\mathrm{local}}$, i.e., the physical error rate of a junction CNOT is $\alpha$ times worse than a local CNOT; and $\ell$, the extra stall ticks for every tick with junction CNOTs. While tuning $\alpha$ is straightforward, we stall the circuit by inserting $\ell$ idle ticks after each tick containing junction CNOTs, during which all qubits idle under the per-tick noise model.

\begin{figure}[t]
    \centering
    \includegraphics[width=\linewidth]{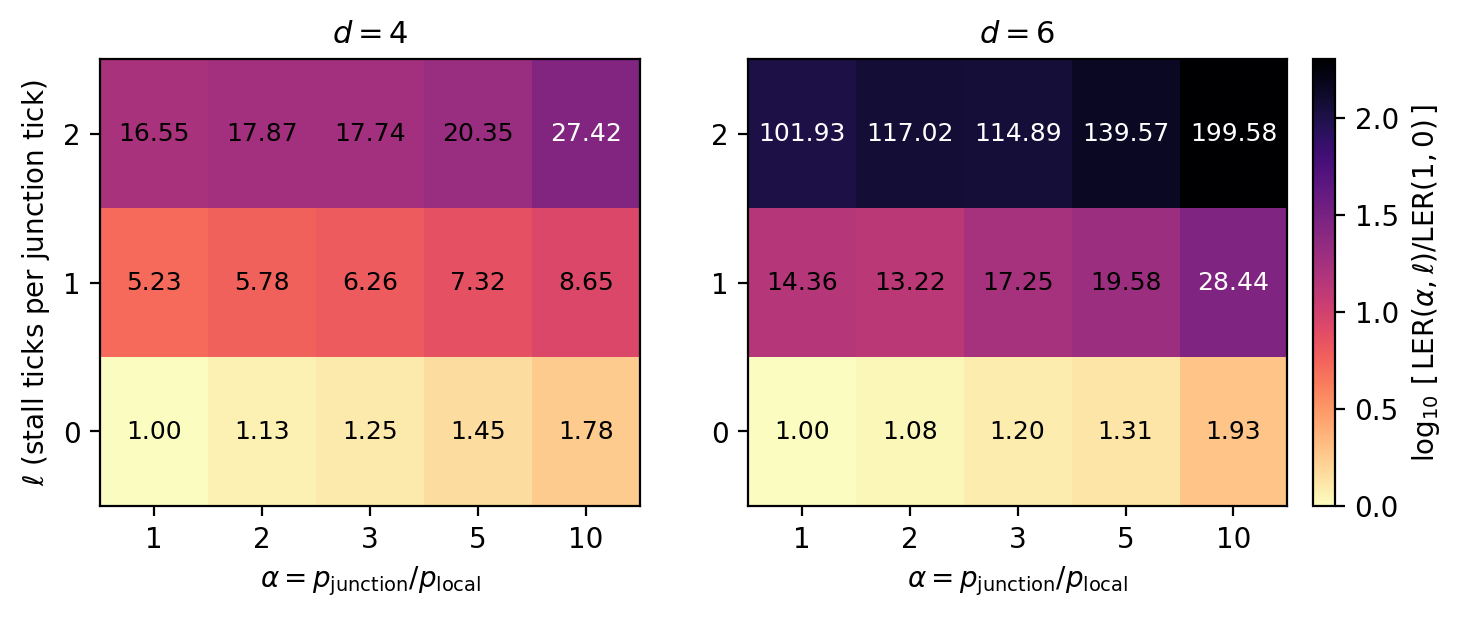}
    \caption{Cost in LER of making the cross-plane CNOTs worse, for the $[[4,2,2]]$ layer code at $\chi=(d,d,d)$, X-basis memory, $p=10^{-3}$. Each cell is the logical error rate at $(\alpha,\ell)$ divided by the same code's own value at $(1,0)$. $\alpha$ scales the depolarizing rate of the junction CNOTs alone; $\ell$ adds that many stalled ticks to every tick holding junction CNOTs, during which every qubit is exposed to the idling noise. Every cell rests on at least $100$ observed logical errors.}
    \label{fig:junction-phase}
\end{figure}

The results of this simulation are reported in \cref{fig:junction-phase}. Each cell is a code's degradation relative to its own value under the default noise model, $\mathrm{LER}(\alpha,\ell)/\mathrm{LER}(1,0)$, measuring how sensitive that code is to the architectural penalty. We sweep the $[[4,2,2]]$ layer code at $\chi=(d,d,d)$ for $d=4$ and $6$ at $p=10^{-3}$. The figure shows X memory, and the values for Z memory are quoted below.

The two axes exhibit unequal effects on the LER. The infidelity of the cross-plane link is relatively mild: a junction CNOT ten times noisier than a local one raises the LER by a factor of only $1.8$ at $d=4$ and $1.9$ at $d=6$, a difference between the two distances that our statistics do not resolve. This is partly because junction CNOTs are a minority of the two-qubit gates: $10.6\%$ at $d=4$ and $6.5\%$ at $d=6$. This ratio is expected, and has been verified, to decrease further as the distance increases, reaching $4.7\%$ at $d=8$: recall from the left panel of \cref{fig:layer-codes} that, by construction, the number of junctions only grows linearly with $d$ while the total CNOT count grows with $d^2$ (the area of the patches). However, it does not follow that the penalty itself shrinks with this ratio. Raising the distance suppresses logical failure by demanding that more faults align, and a larger failure configuration samples more gates, so the shrinking junction share is compensated by the growing number of chances for a junction gate to take part in one. The expected outcome is a penalty that stays roughly flat with distance rather than one that vanishes, as X memory shows. In Z memory the penalty at $\alpha=10$ rises from $1.6\times$ at $d=4$ to $2.5\times$ at $d=6$, which is still modest for a gate ten times noisier, so the mildness is not particular to one memory basis. Thus, in this respect the layer code is not sensitive to the quality of the interconnect that couples its planes, which is the encouraging half of the picture. This implies nothing about the asymptotic behaviour, which larger distances and more expensive simulation would be needed to establish.

In contrast, latency causes a more severe rise in the LER. A single stall tick ($\ell=1$) after every tick that contains a junction CNOT, with every gate error rate left unchanged, costs $5.2\times$ at $d=4$ and $14\times$ at $d=6$ in X memory; two stall ticks cost $17\times$ and $102\times$. The stall penalty thus grows with distance, and at every $\ell$ we sample it dominates the variation along $\alpha$.
In Z memory the same stalls cost $17\times$ ($\ell=1$) and $114\times$ ($\ell=2$) at $d=6$, within $20\%$ of the X values, and $5.1\times$ and $16\times$ at $d=4$. Overall, of the two axes it is latency rather than fidelity that dominates, and thus more effort should go into optimizing the speed of cross-plane operations than into their fidelity.
Note that our model is synchronous: every qubit idles within a stall tick. This is a qualitative approximation of how a machine with slow junction CNOTs would actually schedule an SEC round. A finer-grained model would let the qubits not involved in a junction CNOT carry on with their own part of the round. But each round's duration is set by a critical path on which the stalls lie in sequence. Those qubits therefore finish early and then wait for the ones executing the junction CNOTs before the next round can begin, accumulating idle noise while they wait. Charging every qubit therefore overcounts only those executing the junction CNOTs, at most ${\sim}7\%$ of the register at any stalled tick at $d=6$ and ${\sim}12\%$ at $d=4$. What it does understate is the size of $\ell$. In some implementations a junction CNOT can be one or more orders of magnitude slower than a local two-qubit gate~\cite{bluvstein2024logical}. A realistic machine of that kind therefore sits at a much larger $\ell$ than we simulate, so the penalties in \cref{fig:junction-phase} are a lower bound on its cost. The fidelity parameter $\alpha$ involves no timing assumption and is unaffected by this choice.

\begin{figure*}[t]
    \centering
    \includegraphics[width=0.92\textwidth]{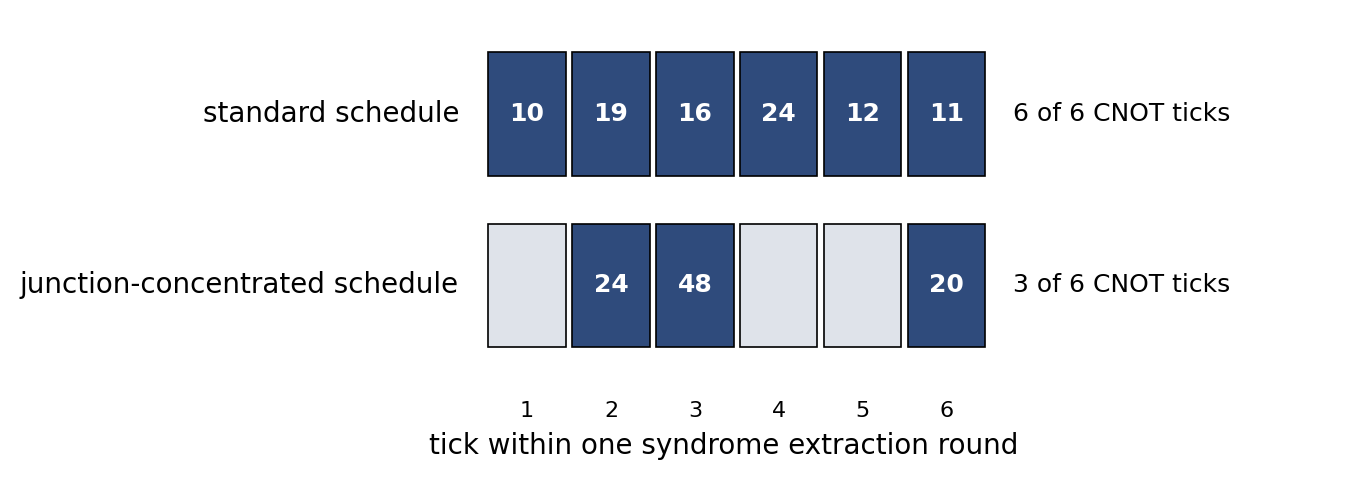}
    \caption{Where the cross-plane CNOTs sit inside one syndrome extraction round of the $[[4,2,2]]$ layer code at $\chi=(6,6,6)$, X-basis memory. Each cell is one of the round's six CNOT-bearing ticks (a round holds one further tick for measurement and reset); dark cells hold at least one junction CNOT, and the number is how many. The $92$ junction CNOTs are $6.5\%$ of the round's $1412$ two-qubit gates, yet under the standard schedule they occupy all six ticks, so an architecture that stalls on a cross-plane operation pays for every tick of the round. Asking the compression step to minimize junction ticks once it has minimized depth and ancilla idle brings six down to three, leaving the depth, the total ancilla idle, the detector count and the number of error mechanisms in the detector error model all unchanged.}
    \label{fig:tick-occupancy}
\end{figure*}

The latency axis therefore gives the SEC scheduler an extra optimization goal. Right now it treats junction and local CNOTs equally, and the junction CNOTs occupy every one of the round's CNOT ticks --- six of six at $\chi=(6,6,6)$ (a round holds one further tick for measurement and reset; \cref{fig:tick-occupancy} shows the six), and five of five at $\chi=(4,4,4)$ --- so a machine that stalls on a cross-plane operation stalls on every CNOT tick of the round. However, since the junction CNOTs of a round form a bipartite graph, by Koenig's theorem the fewest ticks that can hold all of them is its maximum degree, and that degree is two at $d=4$, $6$ and $8$ alike. Up to about a threefold reduction in the number of exposed ticks is therefore in principle available to the scheduler rather than being intrinsic to the construction. Note that the bound counts only the junction CNOTs, so it bounds the achievable benefit rather than constructing it: a real schedule must also place the local CNOTs and respect the determinism and hook-order constraints of \cref{sec:sec-design}. How much of it a schedule actually recovers is measured below.

We therefore give the compression step of \cref{sec:sec-design} a lexicographic objective whose third term is the number of junction ticks, taken up only once depth and ancilla idle have been minimized. That order is deliberate: depth and ancilla idle charge idle noise in every round on any machine, whereas junction ticks charge only if an architecture admits slow cross-plane operations.

At $\chi=(6,6,6)$ the resulting junction-concentrated schedule holds the round at seven ticks unchanged, leaves the total ancilla idle and the detector and mechanism counts of the DEM unchanged, and gathers the junction CNOTs into three ticks from six (\cref{fig:tick-occupancy}). That is one tick short of the two the bound allows, the remaining gap being freedom the optimization has already spent on ancilla idle, which outranks junction count in the objective.

Whether the concentration is worth having depends on the architecture. We compare the concentrated schedule against the standard one on the $[[4,2,2]]$ layer code at $\chi=(6,6,6)$, in both memory bases, at $p=10^{-3}$ and $2\times10^{-3}$, taking the ratio of the concentrated LER to the standard one. Under the default noise model, at $\ell=0$, concentrating costs nothing measurable: the four cells span $0.88$--$1.07$, three are statistically indistinguishable from the standard schedule, and the one that is resolved favours concentrating. Once cross-plane operations are slow the gain is large: all eight cells at $\ell=1$ and $\ell=2$ improve, by $3.3$--$7.0\times$ and $7.2$--$15.2\times$ respectively. Junction concentration is therefore free at this distance, and worth taking whenever the architecture makes cross-plane operations slow.

When concentration is not possible under the standard schedule settings, one can change the lexicographic order of the optimization objective in the scheduler. As an example, at $\chi=(4,4,4)$ the standard schedule leaves no slack: the round is five CNOT ticks deep and the junction CNOTs occupy all five. Promoting the junction count above ancilla idle in the objective brings the number of junction ticks down to three, at the cost of $2.3\%$ in the summed span of the CNOT chains and a doubling of the idle ticks left inside a live ancilla's window. That trade is worth taking as soon as stalls exist: at $p=10^{-3}$ with X memory it costs nothing measurable at $\ell=0$ ($0.99\times$) and returns $1.7\times$ at $\ell=1$ and $2.5\times$ at $\ell=2$.

A layer code therefore makes a narrow demand of its hardware: a small set of cross-plane couplings, fixed by the construction and the same in every round. What matters about them is their latency rather than their fidelity, and how much of that latency the circuit is exposed to is partly a scheduling choice.

\section{Discussion}
\label{sec:discussion}

This paper set out to measure layer codes as a circuit-level quantum memory, under a syndrome extraction schedule designed for them and against surface codes of equal distance, built, noised and decoded the same way. The scheduling side is the part that came out cleanly: on the $[[4,2,2]]$-input codes our circuits lose no distance to hook errors at all. On threshold the layer code is behind: the $[[4,2,2]]$ family reaches $p_{\mathrm{th}}=4.99(6)\times 10^{-3}$ in X memory and $5.06(7)\times10^{-3}$ in Z, against $8.02(6)$ and $8.29(6)\times10^{-3}$ for the unrotated surface code, whose patches are the very ones the layer code is built from, and $7.71(9)$ and $7.73(14)\times10^{-3}$ for the rotated one. On sustained idle robustness the layer code leads, tolerating a $1.7$--$3.0\times$ longer interval between refresh rounds than the rotated surface code at the same distance. It pays for that lead twice over: in control work, running $1.7$ to $5.0$ times more two-qubit gates, measurements and noisy locations per logical qubit per unit time, and in footprint, costing about five times as many physical qubits per logical qubit. We also ran simulations on models inspired by realistic hardware, and the results suggest that the speed rather than the fidelity of the cross-plane two-qubit gates is the bottleneck of the LER. What follows collects the routes that did not work, the limits of what we measured and possible directions for future work.

\subsection{Negative results and limitations}
\label{sec:negative}

Several routes we explored did not produce positive results, and one regime of the construction lies outside what we can characterize at all. We record them here because each one either closes a natural-seeming direction, exposes a structural obstacle that future work on layer codes will meet, or marks the boundary of what the rest of this paper establishes.

\textbf{Concatenated-matching decoders.} The decoder proposed for layer codes in ref.~\cite{williamson2025partialselfcorrectionlayercodes} performs minimum-weight perfect matching independently on each plane and composes the per-plane outcomes through a decoder of the input code. However, its performance guarantee there assumes noiseless syndromes, from which it corrects any error whose energy penalty stays below a constant fraction of the energy barrier. Asking the same construction for a stochastic circuit-level LER is a different question. We therefore extended it to multi-round detector histories with faulty measurements. Furthermore, junction-CNOT faults create hyperedges that no matcher can represent. In our tests, every matching-based variant, from the published per-plane form up to a single global matcher, plateaued far above the rate BP+OSD reaches on the undecomposed model. We therefore also tested a variant in which the per-plane decomposition is retained but each plane is decoded with BP+OSD rather than matching.

We evaluated this decoder on the $[[4,2,2]]$ layer code in the WB24 convention ($318$ data qubits and $316$ stabilizers, $k=2$, $(d_X,d_Z)=(5,6)$), using X-basis memory over $d_Z=6$ rounds. It should be possible to adapt this decoder to our YBW26 convention, but we did not pursue this because it is outperformed by global BP+OSD by orders of magnitude. At $p=10^{-3}$, with CSS-marginal detectors and per-tick idle noise, the per-plane decoding reaches $0.054$ versus $0.0005$ for global decoding of the identical DEM in $2\times10^{4}$ shots. Our variant does not work: the per-plane decomposition itself is what costs the two orders of magnitude. This is an empirical ceiling on it rather than a theorem, and it is why every result in this paper uses global decoding of the raw DEM.

We add two asides. First, Tesseract~\cite{beni2025tesseractdecoder}, a near-most-likely-error decoder, was orders of magnitude slower than global BP+OSD on these hyperedge-dense models at every beam width we tried. Second, we did not adapt the cluster decoder of ref.~\cite{layercodepy2510} to the circuit-level noise model: it decodes the whole code rather than plane by plane, but its threshold proof, like the guarantee above, assumes noiseless syndromes.

\textbf{SEC scheduling routes.} A few scheduling ideas were rejected before the design of \cref{sec:sec-design} settled; here we mention two instructive examples. The first was to extract reusable local CNOT-order templates from LRC's metric-best schedules. It fails because those schedules carry no low-dimensional structure: stabilizers of the same geometric type receive mutually inconsistent optimal orders. The second was to fix the round depth first and repair the determinism condition afterwards, the opposite order to \cref{sec:sec-design}. Repair at fixed depth plateaus with a third to a half of the check pairs still violating the parity condition of \cref{app:sec-algorithm}. Our scheduler therefore imposes that condition on its search rather than repairing after the fact.

\textbf{Storage efficiency.} We asked whether some input code makes a layer code memory cheaper, in physical qubits per logical qubit at equal distances, than rotated surface code patches. A closed-form prescreen (\cref{app:storage}) reduces the question to the ratio $R_{\mathrm{opt}}\approx2/(\alpha_X\alpha_Z r)$, where $r$ is the input encoding rate and $\alpha_X=d_X/\chi_Z$, $\alpha_Z=d_Z/\chi_X$ measure how efficiently the construction converts patch dimension into output distance; a storage win requires $R_{\mathrm{opt}}<1$ before decoder effects are even considered. 

No candidate we screened clears that condition: the most favourable, a layer code built from a bivariate-bicycle $[[18,8,2]]$ input~\cite{bravyi2024high}, reaches only $R_{\mathrm{opt}}\approx1.1$, and that is already an optimistic bound. Since $R_{\mathrm{opt}}<1$ demands $\alpha_X\alpha_Z r>2$ while neither amplification factor exceeded the input code distance in any family we screened, only higher-distance inputs remain; for $d\geq5$, low-check-weight, non-trivial-rate LDPC inputs do not exist at tractable sizes in the families we scanned (bivariate-bicycle, lifted-product~\cite{panteleev2022almost}, and asymmetric high-$d_X$ constructions), with the input check weight as the binding obstruction. The one $d=4$ candidate we could build and measure, a two-dimensional yoked parity-check input~\cite{gidney2025yoked}, bears the prescreen out: its layer code decodes cleanly yet costs about $16\times$ more qubits per logical qubit than a rotated surface patch of equal or better LER. The same accounting sets a floor against the input itself: at the same distance a layer code costs at least $4\times$ more qubits per logical qubit than the code it is built from. Storage efficiency is therefore not what this construction offers; its value lies in locality, bounded check weight, and the idle robustness of \cref{sec:idle}. We emphasize that the scope of this negative result is narrow: it applies to the families we screened, not to layer codes in general, and no $d\geq5$ data point at the same LER exists on either side.

\textbf{Quantum Tanner inputs.} The optimal parameters proved for layer codes are established with quantum Tanner codes as the input family~\cite{layercodes2309}, whereas every layer code studied above is built from a small input. We did apply the construction to that class, using the explicit $[[36,8,3]]$, $[[54,11,4]]$ and $[[72,14,4]]$ instances of ref.~\cite{radebold2025explicit}. They give layer codes on $15780$, $24774$ and $21544$ physical qubits, encoding $8$, $11$ and $14$ logical qubits, both counts exact. Their distances we can only bound from above, at $(d_X,d_Z)\le(58,30)$, $(56,42)$ and $(24,52)$ respectively, and those bounds need not be tight; certifying one from below is, by our estimate, orders of magnitude out of reach. An exact characterization of a layer code built from a quantum Tanner input therefore remains open. What would change that is not more computation but a theorem making explicit the constant in the construction's distance bound.

\subsection{Outlook}
As for future work, since our scheduler uses no layer code structure beyond the lattice metric behind its hook rule, a natural continuation is a code-agnostic syndrome extraction compiler that treats the ancilla live window and certified hook structure as first-class objectives alongside the circuit depth that general compilers already optimize~\cite{zhang2026optimal}.

A second direction is logical operations, of which this paper measures none: which ones a layer code admits beyond storage, and how they compare with the lattice surgery a surface code would use~\cite{latticesurgery1111}, are open. Transversal CNOT is the natural first example. Beyond the LER, an axis worth measuring is throughput under an explicit connectivity and workspace budget, where one $k$-logical block performs $k$ paired CNOTs at once.

A third direction is dimension: layer codes have been generalized to $D=4$ and $5$ by color routing~\cite{yuan2026colorrouting}, which saturates the BPT bound in those dimensions with good qLDPC input. Whether the better parameters carry over to a circuit-level advantage is open, since the construction's check weight $3(D-1)$ and qubit degree $2D$ force a deeper extraction round, and depth is the cost that dominates the threshold gap we measure in three dimensions.

\begin{acknowledgments}
The authors thank Dominic Williamson, Shouzhen Gu, Zhiyang He, B\'alint Koczor and Simon Benjamin for fruitful and inspiring discussions. Z.C. acknowledges support from the EPSRC Quantum Technologies Career Acceleration Fellowship (UKRI1226). The authors acknowledge the use of the University of Oxford Advanced Research Computing (ARC) facility in carrying out this work: \url{http://dx.doi.org/10.5281/zenodo.22558}.
\end{acknowledgments}

\section*{Author contributions}

Z.S. and Z.C. conceived the study. Z.S. designed the syndrome extraction scheduler, benchmarked the performance of the circuits and wrote the original draft of the manuscript. Z.C. supervised the project. Both authors reviewed and edited the manuscript. Large language models were used to implement the simulation and analysis code, to polish parts of the text, and to check the bibliography.

\section*{Code and data availability}

The code that generates the syndrome extraction circuits, the decoder, the analysis scripts, and the circuits and sampled logical error rates behind every figure and table in this paper are available on Zenodo~\cite{layercodes2026artifact}.

\bibliography{refs}

\appendix
\crefalias{section}{appendix}

\section{Numerical methods and code parameters}
\label{app:methods}

\subsection{Sampling and statistics}

All circuits are built and sampled with Stim~\cite{gidney2021stim} (version 1.15.0). The threshold scans of \cref{sec:memory-ler} and the hardware grids of \cref{sec:hardware} collect their statistics through Sinter (version 1.15.0) running the decoder described in the next subsection; every other simulation runs the same decoder on samples drawn with fixed seeds from Stim's detector sampler. With the exceptions noted below, every decoding problem in this paper is the undecomposed DEM of the noisy circuit, decoded globally: mechanisms are never decomposed into graphlike pairs, and the model is never split by plane. The exceptions are the per-plane decoder evaluated in \cref{sec:negative}, for which that split is the object of study, and the code-capacity curves of \cref{fig:sec-benchmark}, which have no circuit and are decoded directly on the code's parity checks. The windowed re-analysis of \cref{sec:idle} decodes the same undecomposed model, but commits its correction in a sliding window rather than all at once.

Each point (code, memory basis, physical error rate) is sampled to a shot cap or a logical-error cap, whichever comes first. The threshold scans of \cref{sec:memory-ler} use $4\times10^{5}$ shots or $4\times10^{3}$ errors per point, on code-specific grids of seven or eight rates bracketing the crossing region; the two circuits of \cref{fig:sec-ler-lrc} use $3\times10^{5}$ or $3\times10^{3}$, both sampled in the same run; the noise-model benchmark of \cref{fig:sec-benchmark} uses $2\times10^{5}$ or $2\times10^{3}$, raised to $1.5\times10^{7}$ shots on the code-capacity floor; the hardware grids of \cref{sec:hardware} use $2\times10^{5}$ shots or $2\times10^{3}$ errors per cell, topped up to at least $10^{2}$ errors on sparse cells and so to ${\sim}10^{7}$ shots, with the schedule comparison of \cref{fig:tick-occupancy} run to $3\times10^{7}$; and the sustained-memory grids of \cref{sec:idle} use code-dependent budgets, from $5\times10^{4}$ shots or $400$ errors per point to $3\times10^{5}$ or $800$, with the zero-idle anchor points sampled to $3\times10^{7}$ shots; and the windowed re-analysis and the single-window column of \cref{sec:idle} use $10^{5}$ shots per point.

Error bars on sampled LER are Wilson $95\%$ intervals. Independently sampled circuits are compared with an unpaired two-proportion $z$ test; the per-shot paired decoder comparison of \cref{app:lsd-osd} uses McNemar's exact test.

\subsection{Decoder configuration}

Every circuit-level decode in this paper uses the same implementation and settings, except for the choice between BP+LSD and BP+OSD. Belief propagation runs min-sum updates with scaling factor $1.0$, a parallel schedule and at most $30$ iterations. Its priors are the mechanism probabilities of the DEM, so the decoder is automatically matched to whichever noise model generated the circuit. Post-processing is a combination sweep of order $3$: the LSD form for BP+LSD, the OSD form for BP+OSD. Both come from \texttt{ldpc} 2.4.1~\cite{roffe2020decoding,hillmann2024lsd}. The code-capacity curves of \cref{fig:sec-benchmark} use the same library decoder with identical settings, but one instance per Pauli sector. It acts on the code's parity-check matrix with a uniform prior at the sampled error rate, not on a DEM. A shot counts as a logical error when the decoded value of any of the $k$ observables differs from the recorded one, normalized as defined in \cref{sec:decoding}.

\subsection{Code and circuit parameters}

\Cref{tab:arms-layer} lists every YBW26 layer code used in this paper and \cref{tab:arms-surface} the surface codes. It is worth noting that the number of syndrome-extraction rounds is keyed to the opposite-basis distance (\cref{sec:background}), so the two memory experiments of an asymmetric code run for different numbers of rounds. Their rounds are built from the same CNOT set at the same depth, so the tick and CNOT counts listed here are basis independent, but each basis is scheduled by its own run of the search of \cref{app:sec-algorithm} and the two place a given CNOT in different ticks. A tick is a time slice, and on every circuit exactly one tick per round carries no CNOT, so the CNOT depth quoted in \cref{sec:sec-design} is one less than the tick count listed here.

\begin{table*}[t]
\begin{ruledtabular}
\begin{tabular}{llcccccc}
family & $\chi=(\chi_X,\chi_Q,\chi_Z)$ & $k$ & $n_{\mathrm{data}}$ & $n$ & $(d_X,d_Z)$ & ticks/round & CNOTs/round \\
\hline
$[[4,2,2]]$ & $(4,4,4)$ & 2 & 148 & 294 & $(4,4)$ & 6 & 564 \\
 & $(5,5,5)$ & 2 & 244 & 486 & $(5,5)$ & 7 & 940 \\
 & $(6,6,6)$ & 2 & 364 & 726 & $(6,6)$ & 7 & 1412 \\
 & $(7,7,7)$ & 2 & 508 & 1014 & $(7,7)$ & 7 & 1980 \\
 & $(8,8,8)$ & 2 & 676 & 1350 & $(8,8)$ & 7 & 2644 \\
\hline
$[[6,4,2]]$ & $(4,6,4)$ & 4 & 226 & 448 & $(4,4)$ & 7 & 866 \\
 & $(5,6,5)$ & 4 & 344 & 684 & $(5,5)$ & 7 & 1334 \\
 & $(6,6,6)$ & 4 & 486 & 968 & $(6,6)$ & 7 & 1898 \\
 & $(7,6,7)$ & 4 & 652 & 1300 & $(7,7)$ & 7 & 2558 \\
 & $(8,6,8)$ & 4 & 842 & 1680 & $(8,8)$ & 7 & 3314 \\
\hline
Steane $[[7,1,3]]$ & $(3,4,3)$ & 1 & 193 & 385 & $(7,6)$ & 8 & 784 \\
 & $(3,7,3)$ & 1 & 283 & 565 & $(8,7)$ & 8 & 1116 \\
 & $(4,4,4)$ & 1 & 319 & 637 & $(9,8)$ & 9 & 1296 \\
 & $(4,7,4)$ & 1 & 445 & 889 & $(10,9)$ & 8 & 1772 \\
 & $(5,4,5)$ & 1 & 473 & 945 & $(11,10)$ & 9 & 1920 \\
\hline
Shor $[[9,1,3]]$ & $(2,6,4)$ & 1 & 271 & 541 & $(4,8)$ & 9 & 1008 \\
 & $(3,6,5)$ & 1 & 467 & 933 & $(5,11)$ & 8 & 1788 \\
 & $(4,6,6)$ & 1 & 699 & 1397 & $(6,14)$ & 10 & 2712 \\
\end{tabular}
\end{ruledtabular}
\caption{The YBW26 layer codes used in this paper. $n_{\mathrm{data}}$ is the number of data qubits and $n$ the total physical qubit count of the syndrome extraction circuit, data plus one ancilla per stabilizer. Distances are code-capacity distances of the layer code, certified as described in \cref{sec:memory-ler}. The number of syndrome extraction rounds is the opposite-basis distance, so a code with $d_X\neq d_Z$ runs for different numbers of rounds in the two memory experiments; both the tick count and the CNOT count per round are basis independent. All rows use the SEC of \cref{sec:sec-design}, and the counts are taken from the circuits we simulate.}
\label{tab:arms-layer}
\end{table*}

\begin{table}[t]
\begin{ruledtabular}
\begin{tabular}{lccccc}
code & $d$ & $n$ & rounds & ticks/round & CNOTs/round \\
\hline
unrotated & 3 & 25 & 3 & 5 & 40 \\
 & 4 & 49 & 4 & 5 & 84 \\
 & 5 & 81 & 5 & 5 & 144 \\
 & 6 & 121 & 6 & 5 & 220 \\
 & 7 & 169 & 7 & 5 & 312 \\
 & 8 & 225 & 8 & 5 & 420 \\
\hline
rotated & 3 & 17 & 3 & 5 & 24 \\
 & 4 & 31 & 4 & 5 & 48 \\
 & 5 & 49 & 5 & 5 & 80 \\
 & 6 & 71 & 6 & 5 & 120 \\
 & 7 & 97 & 7 & 5 & 168 \\
 & 8 & 127 & 8 & 5 & 224 \\
\end{tabular}
\end{ruledtabular}
\caption{The surface codes, simulated in the same way as the layer codes. Both families have $k=1$ and $d_X=d_Z=d$, so both memory experiments run for $d$ rounds. $n$ counts data plus ancilla qubits. The rotated circuits are Stim's generated circuits with the basis-conjugating Hadamards absorbed into native-basis resets and measurements (\cref{sec:memory-ler}); the unrotated ones come from our own builder with the balanced alternating CX schedule of ref.~\cite{orourke2025comparepair}, whose consequences are controlled for there.}
\label{tab:arms-surface}
\end{table}

\subsection{The simulated circuits}

Phase one of the scheduler (\cref{app:sec-algorithm}) is deterministic, but phase two is an anytime constraint-programming search under a wall-clock budget, and it runs on several worker threads. A rebuild therefore returns a different schedule of the same quality even on the same machine under the same budget: the same proven-optimal depth, the same ancilla idle to within $1.5\times10^{-4}$ of a proven bound, a different placement of individual gates. Every layer code number reported here refers to the specific circuits we generated rather than to the procedure that produced them, and those circuits are part of the public release.

For the per-round idle model of \cref{sec:memory-ler} we use a second version of each circuit, in which every ancilla is reset at the start of the round and measured at its end, instead of immediately before its first CNOT and after its last. Charging idle noise once per round requires a well-defined round boundary, and the charge does not depend on where within the round an ancilla waits. Both versions are built from the same schedule, so their CNOTs coincide tick for tick on every code.

\section{The syndrome extraction scheduler}
\label{app:sec-algorithm}

This appendix states the scheduler of \cref{sec:sec-design} as implemented. The input is the stabilizers of the layer code, each with its data support and ancilla, together with the lattice coordinates and, for each type, the least-weight of the $k$ cleaned logical representatives. The output is a single syndrome extraction round, as an ordered list of ticks of disjoint CNOTs. Write $\mathrm{supp}(s)$ for the data support of stabilizer $s$, $a(s)$ for its ancilla, and $w_s=|\mathrm{supp}(s)|$. Write $D(q)$ for the L1 lattice distance from data qubit $q$ to the nearest qubit of that representative of the same type. CNOT directions follow the stabilizer type rather than the memory basis. An X-type stabilizer contributes $\mathrm{CNOT}(a(s),q)$, a Z-type one $\mathrm{CNOT}(q,a(s))$.

As an overview, phase one of the scheduler (\cref{alg:phase1}) enforces the hook-error constraint: it reduces each stabilizer to the CNOT orders that satisfy it, and places one of them to give a valid schedule. Phase two (\cref{alg:phase2}) determines the timing: it assigns every CNOT a tick and, in the same optimization, the order each stabilizer uses. Neither calls a decoder.

\begin{algorithm}[t]
\caption{Phase one: hook-constrained placement, run separately for each stabilizer type $\beta\in\{X,Z\}$}
\label{alg:phase1}
\KwIn{stabilizers $S_\beta$; distances $D(\cdot)$}
\KwOut{candidate order sets $\{\Pi(s)\}_{s\in S_\beta}$; a feasible tick list $\mathcal{T}_\beta$}
\ForEach{$s\in S_\beta$}{
  $\Pi(s)\gets$ the residual-admissible orderings $\pi$ of $\mathrm{supp}(s)$ whose last qubit attains $\max_{q\in\mathrm{supp}(s)}D(q)$\;
  $\pi(s)\gets$ the least element of $\Pi(s)$ in lexicographic order\;
}
$\mathcal{T}_\beta\gets$ an empty tick list, extended on demand\;
\ForEach{$s\in S_\beta$ in index order}{
  $t\gets-1$\;
  \ForEach{$q\in\pi(s)$ in order}{
    $t\gets\min\{t'>t:\ a(s)$ and $q$ are both free in $\mathcal{T}_\beta[t']\}$\;
    add the CNOT between $a(s)$ and $q$ to $\mathcal{T}_\beta[t]$\;
  }
}
\end{algorithm}

$\Pi(s)$ combines the geometric rule of \cref{sec:sec-design} with an admissibility condition inherited from LRC. The rule makes the hook-error decision, requiring the last CNOT to act on a data qubit farthest from the logical representative, where the residual error left by an ancilla fault is least likely to complete into a logical operator. Admissibility keeps only orderings whose every CNOT suffix is a residual that LRC's left-right construction can produce. That test is purely combinatorial; it displaces the lexicographically first rule-compliant ordering for a minority of stabilizers, and no stabilizer is left without an admissible ordering. Selection among the survivors uses no decoder metric, and the greedy pass places every stabilizer at its first candidate, giving phase two a feasible starting point; the survivor each stabilizer finally uses is settled there.

\begin{algorithm}[t]
\caption{Phase two: depth compression and idle minimization}
\label{alg:phase2}
\KwIn{tick lists $\mathcal{T}_X,\mathcal{T}_Z$; candidate order sets $\{\Pi(s)\}$; time budget $B$}
\KwOut{merged tick list $\mathcal{T}$}
introduce an integer variable $t_e$ for every CNOT $e$\;
\ForEach{stabilizer $s$}{
  constrain the order the $t_e$ induce on $\mathrm{supp}(s)$ to lie in $\Pi(s)$\;
}
constrain the $t_e$ of CNOTs sharing a qubit to be pairwise distinct\;
\ForEach{pair of an X-type stabilizer $s$ and a Z-type stabilizer $s'$}{
  constrain the number of shared data qubits $q\in\mathrm{supp}(s)\cap\mathrm{supp}(s')$ whose Z-CNOT precedes its X-CNOT to be even\;
}
$m^{*}\gets$ the minimum of $\max_e t_e$ subject to the constraints above\;
constrain $\max_e t_e\le m^{*}$\;
minimize $\sum_s\left(\max_{e\in s}t_e-\min_{e\in s}t_e\right)$, tie-broken by $\sum_e t_e$, warm-started from the makespan solution\;
\Return the best schedule found within $B$, falling back to the depth-optimal solution of the makespan step, or to a deterministic greedy merge if no solver is available\;
\end{algorithm}

Phase two performs the final tick assignment. It merges $\mathcal{T}_X$ and $\mathcal{T}_Z$ into the single tick list $\mathcal{T}$, choosing each stabilizer's CNOT order from $\Pi(s)$ as part of the same optimization, and optimizes lexicographically. The first level compresses depth and then freezes it for the following idle reduction. The second level minimizes the total ancilla idle sitting between the first and last CNOT of each chain. The concentrated variant of \cref{sec:hardware} inserts the junction-tick count as an optional third level above the tie-break. Placing it above the ancilla-idle level instead trades idle for junction ticks concentration. Resets and measurements are then staggered against the chains. Each ancilla is reset in the segment just before its own first CNOT, and measured in the segment just after its last.

Three features of the model deserve comment. The cross-type constraint is what makes the merge safe. An $X$- and a $Z$-type stabilizer of a CSS code share an even number of data qubits, exactly two on every sharing pair of the codes here. Each shared qubit carries two CNOTs, one out of the $X$-type ancilla $a$ and one into the $Z$-type ancilla $b$, and swapping their order changes the round by a $\mathtt{CNOT}(a,b)$. Swapping both shared qubits therefore changes nothing, the two factors cancelling; swapping one leaves a CNOT between the ancillas, so neither ancilla's measurement carries its own check value and the first-round detectors are no longer deterministic. Even parity makes the two circuits equivalent, which is Proposition~1 of ref.~\cite{zhang2026optimal}; ordering every X-CNOT before every Z-CNOT is its zero-inversion special case, so the constraint is always satisfiable.\footnote{Their statement uses an $X$-basis ancilla for both check types, with $\mathtt{CZ}$ for the $Z$-checks. Conjugating that ancilla by a Hadamard turns $\mathtt{CZ}$ into our $\mathtt{CX}(q,a)$ without moving any gate in time, so their inversion number across a check pair's shared qubits becomes the count of those qubits whose $Z$-CNOT precedes its $X$-CNOT.} The idle objective is expressed as a sum of CNOT chain spans, which equals that idle count up to the constant $\sum_s(w_s-1)$ because each chain is monotone. And only that idle is worth minimizing, because noise on an ancilla before its reset or after its measurement cannot reach the syndrome, so the staggering removes the rest without touching the CNOT schedule, the tick count or the detector structure.

Phase two runs as two successive time-limited optimizations. The first minimizes depth. The second carries the remaining levels, i.e., the idle objective with the junction term and the tie-break below it. Each reports the best schedule found within its share of the time budget $B$, together with whether optimality was proved. If the depth step does not prove optimality, it fixes the depth at the best value it found, which may then exceed the optimum. On the codes of this paper the minimal depth is proved on every circuit, in anything from a second to about a minute, so the rest of the budget goes to the idle objective. The final tie-break prefers earlier placements. It leaves the total number of noise locations unchanged and exists only to shrink the degeneracy among equally optimal schedules. The solver runs with four parallel search threads, so its search is not reproducible from run to run. The value of the idle objective is nonetheless reproducible: on every circuit we use, the second optimization ends within $1.5\times10^{-4}$ of its own proven lower bound, so independent runs agree on the quantity being minimized. What they can differ in is where individual gates sit within that optimum, which the objective does not see. This is why every result in this paper refers to the published circuits rather than to a rerun of the scheduler (\cref{app:methods}).

\section{BP+LSD versus BP+OSD on layer code circuits}
\label{app:lsd-osd}

Two datasets quantify the accuracy gap between our default decoder, BP+LSD, and the accuracy reference, BP+OSD.

First, in the noise-model benchmark of \cref{sec:sec-design} we decode the $[[4,2,2]]$ family at $\chi=(d,d,d)$ with both decoders on the same seeds, under the default noise model, at the same physical error rates; the noisy circuits are the same as those behind \cref{fig:memory-threshold}. The gap at the same $p$, at $d=6$:

\begin{table}[h]
\begin{ruledtabular}
\begin{tabular}{lcc}
$p$ & LSD$-$OSD (X memory) & LSD$-$OSD (Z memory) \\
\hline
$3.0\times10^{-3}$ & $+0.03$ pp & $+0.01$ pp \\
$4.3\times10^{-3}$ & $+0.14$ pp & $+0.16$ pp \\
$5.0\times10^{-3}$ & $+0.37$ pp & $+0.39$ pp \\
$6.0\times10^{-3}$ & $+0.89$ pp & $+0.54$ pp \\
\end{tabular}
\end{ruledtabular}
\caption{Logical-error-rate difference between BP+LSD and BP+OSD at the same physical error rate, in percentage points, on the $[[4,2,2]]$ family at $d=6$ (our SEC and noise conventions; the two decoders decode the same seeds, paired up to the shorter shot count where the adaptive stopping differs).}
\label{tab:lsd-osd}
\end{table}

BP+LSD is systematically the pessimistic decoder, and the gap grows with $p$ into the error-rich regime. Extending the comparison to the whole family ($d=4$ to $8$, both bases, thirteen physical error rates) gives the same sign at every statistically resolved point, with gaps from $+0.01$ to $+4.3$ pp. The relative excess grows with distance: taking the per-point ratio over all pairs with at least $10^{2}$ errors under both decoders, its median rises from $1.17$ at $d=4$ to $1.36$ at $d=8$. This is a pointwise comparison of LER at fixed $p$, made before any threshold fitting, and is therefore a clean read of the decoder bias.

Second, a per-shot paired calibration, in which both decoders decode identical shots and discordant outcomes are tested with McNemar's exact test, was run on our $d=6$ circuits (CSS-marginal detectors, X and Z memory, default noise model). It locates the rate above which the choice of decoder matters at all. At the operating point $p=10^{-3}$ the two decoders are indistinguishable: over $2.2\times10^{5}$ shots across the two bases, two discordant shots occur, bounding the LER difference by $10^{-5}$. The difference is still unresolved at $p=2.6\times10^{-3}$, where it measures $+0.01$ pp (X) and $+0.03$ pp (Z) ($p_{\mathrm{McNemar}}=0.50$ and $0.18$), and becomes significant only from $p\approx3.6\times10^{-3}$, reaching $+0.23$ pp (X) and $+0.34$ pp (Z) at $p=5\times10^{-3}$ ($p_{\mathrm{McNemar}}=1.0\times10^{-2}$ and $5.3\times10^{-5}$). BP+LSD is on the high side at every sampled point.

Both datasets agree on the direction: on layer code circuits BP+LSD errs on the pessimistic side, with the excess growing with $p$ across the fitting window. LSD-fitted LER curves therefore run slightly high, and the fitted thresholds slightly low; the thresholds quoted in \cref{sec:memory-ler} are, if anything, slight under-estimates.

\section{Storage-efficiency prescreen}
\label{app:storage}

We quantify qubit overhead by $\omega=n_{\mathrm{phys}}/(k\,d^{2})$, the number of physical qubits (data plus ancilla) per logical qubit at distance $d$; a rotated surface code patch has $\omega_{\mathrm{surf}}=2$. For a layer code with input parameters $[[n,k,d_{\mathrm{in}}]]$, rate $r=k/n$, and $n_X$ ($n_Z$) input checks, every patch is an unrotated surface code patch, so at leading order in the patch dimensions $n_{\mathrm{phys}}\approx4\left(n\,\chi_X\chi_Z+n_X\,\chi_Q\chi_Z+n_Z\,\chi_X\chi_Q\right)=4s\,n\,\chi_X\chi_Z$, where $s\ge1$ collects the check-patch overhead, while the output distances are $d_X=\alpha_X\chi_Z$ and $d_Z=\alpha_Z\chi_X$ for amplification factors $\alpha_X,\alpha_Z$ set by the construction; across the families we screened neither value exceeded the input code distance. The two distance-carrying dimensions are independent parameters, so matching the surface code at a common distance $d$ means setting $\chi_Z=d/\alpha_X$ and $\chi_X=d/\alpha_Z$; the patch dimensions then cancel in the ratio, leaving the closed-form prescreen
\begin{equation}
R_{\mathrm{opt}}=\frac{\omega_{\mathrm{layer}}}{\omega_{\mathrm{surf}}}=\frac{2s}{\alpha_X\,\alpha_Z\,r}\,,
\end{equation}
which we evaluate optimistically at $s=1$, dropping the check patches entirely. $R_{\mathrm{opt}}<1$ is necessary but not sufficient for a storage win: the operational verdict is taken at equal LER, $R_{\mathrm{LER}}\approx R_{\mathrm{opt}}\kappa^{2}$, where $\kappa$ rescales distances for the difference in thresholds and prefactors and must be measured. The same accounting yields $\omega_{\mathrm{layer}}/\omega_{\mathrm{input}}=4d_{\mathrm{in}}^{2}/(\alpha_X\alpha_Z)\ge4$: a layer code never stores more cheaply than its own input code; what it buys is 3D locality and check weight at most 6.

The YBW26 construction assigns the qubits and checks of the input code to coordinates along the patch dimensions through a colouring map $\eta$, so that non-conflicting ones may share a coordinate: the trivial (identity) $\eta$ gives each its own coordinate, whereas a nontrivial $\eta$ compresses the patch dimensions toward the chromatic minima of the input's conflict graphs~\cite{yuan2026quantumweightreductionlayer}. Under the trivial-$\eta$ convention the prescreen fails for every input we screened, namely $[[4,2,2]]$, $[[6,4,2]]$, Steane $[[7,1,3]]$, Shor $[[9,1,3]]$ and the $[[15,7,3]]$ quantum Hamming code, which give $R_{\mathrm{opt}}$ between $2.2$ and $4.3$ at $s=1$, with the output distances of all but the $[[15,7,3]]$ row certified by integer programming; that row's $d_X$ and $d_Z$ are weight-4 witnesses, and a smaller true distance would only raise its $R_{\mathrm{opt}}$. The mechanism is structural: trivial $\eta$ forces each patch dimension to be at least the corresponding input multiplicity ($\chi_X\ge n_X$, $\chi_Q\ge n$, $\chi_Z\ge n_Z$), so for any input whose check count is comparable to its distance the amplification collapses to $\alpha\approx1$ (for the Shor input, $n_Z=6\ge d_X=6$ gives $\alpha_X=1$ exactly).

Chromatic compression (nontrivial $\eta$) removes the multiplicity floor, since the patch dimensions then scale with those chromatic numbers rather than with the input's size, and it genuinely reopens the search; the ceiling moves but does not clear the bar: the most favourable candidate our scans produced, a layer code built from a bivariate-bicycle $[[18,8,2]]$ input~\cite{bravyi2024high}, reaches only $R_{\mathrm{opt}}\approx1.1$, and even that is an optimistic bound whose amplification factors we did not certify. Clearing $R_{\mathrm{opt}}<1$ therefore needs a $d\ge5$ input with low check weight and non-trivial rate, since the check weight enters the overhead quadratically; scans of bivariate-bicycle, lifted-product~\cite{panteleev2022almost}, and asymmetric high-$d_X$ constructions found none at tractable sizes. The storage question thus remains open: the necessary-condition failures above cover the families scanned, and no $d\ge5$ comparison at equal LER exists on either side.

\end{document}